\newcommand{\w}{\omega}
\newcommand{\TK}{T_{\rm K}}
\newcommand{\Tc}{T_{\rm c}}
\newcommand{\Jc}{J_{\rm c}}
\newcommand{\rmax}{r_{\rm max}}
\newcommand{\Simp}{S_{\rm imp}}

\newcommand{\al}{a_{\rm l}}

\newcommand{\epsfb} {\varepsilon_0}     
\newcommand{\epsfc} {\varepsilon_c}     
\newcommand{\epsf}  {\varepsilon}       

\newcommand{\ybco}{YBa$_2$Cu$_3$O$_{y}$}
\newcommand{\bscco}{Bi$_2$Sr$_2$CaCu$_2$O$_{8+\delta}$}

\newcommand{\eqref}[1]{{(\ref{#1})}}

\documentclass[12pt]{iopart}
\usepackage{iopams}
\usepackage{tabularx,graphicx}
\usepackage{color}

\begin{document}

\title[Kondo Impurities in Graphene]{The Physics of Kondo Impurities in Graphene}

\author{Lars Fritz}
\address{Institut f\"ur Theoretische Physik, Universit\"at zu K\"oln, 50937 K\"oln, Germany}
\ead{lsfritz@thp.uni-koeln.de}
\author{Matthias Vojta}
\address{Institut f\"ur Theoretische Physik, Technische Universit\"at Dresden, 01062 Dresden, Germany}
\ead{matthias.vojta@tu-dresden.de}

\begin{abstract}
This article summarizes our understanding of the Kondo effect in graphene, primarily from
a theoretical perspective. We shall describe different ways to create magnetic moments in
graphene, either by adatom deposition or via defects. For dilute moments, the theoretical
description is in terms of effective Anderson or Kondo impurity models coupled to
graphene's Dirac electrons. We shall discuss in detail the physics of these models,
including their quantum phase transitions and the effect of carrier doping, and confront
this with existing experimental data. Finally, we point out connections to other quantum
impurity problems, e.g., in unconventional superconductors, topological insulators, and
quantum spin liquids.
\end{abstract}

\maketitle


\section{Introduction}
\label{sec:intro}

The low-temperature behaviour of dilute magnetic impurities in metals, known as the Kondo
effect, is a prime example of electron-correlation physics. The impurity's magnetic
moment is screened below a temperature $\TK$ by the formation of a many-body singlet with
the conduction-electron bath. The Kondo temperature $\TK$ itself depends in a
non-analytic fashion on the Kondo coupling and the bath density of states, signalling the
breakdown of perturbation theory. Starting with Kondo's work in the 1960s \cite{kondo},
we now have an essentially complete set of theoretical descriptions of the Kondo effect
available \cite{hewson}, and agreement between theory and experiment has been established
on a quantitative level.

The understanding of the Kondo effect in metals has prompted to investigate similar impurity
physics in other settings, with the overarching goal to employ impurities as local
probes of the host's properties .
This article is devoted to a particularly interesting and timely case, namely the Kondo
effect in graphene. Here, the impurity spin interacts with the Dirac fermions of the
two-dimensional (2d) sheet of carbon atoms~\cite{novo1,novo2,neto_rmp,sarma_rmp}. For a
local magnetic impurity in charge-neutral graphene, this results in the Kondo effect
being qualitatively different from that in conventional metals, because the bath density
of states now vanishes at the Fermi level: Kondo screening is suppressed at small Kondo
couplings, and a non-trivial impurity quantum phase transition \cite{mv_rev} between an unscreened and a
screened impurity spin obtains.
In fact, this phase transition has first been discussed for magnetic impurities in
unconventional superconductors, where field theories and numerical solutions for the
resulting pseudogap Kondo problem have been worked out.
Graphene offers the attractive possibility of tuning the chemical potential relative to
the Dirac point, such that the crossover between pseudogap and conventional Kondo physics
can be accessed in detail. Furthermore, the 2d nature of graphene naturally allows one to
employ scanning-probe techniques to locally study impurity physics.

In this article we review the theoretical understanding of the Kondo effect in graphene,
together with the current status of experiments. As we will discuss, clear-cut
experimental verifications of some of the exciting theoretical ideas are lacking, and we
contemplate on possible sources of complications. We also highlight theoretical
connections between the Kondo effect in graphene and other quantum impurity problems,
such as impurities on the surface of topological insulators.

\subsection{Outline}

The body of this article is organized as follows:
In Sec.~\ref{sec:imps} we introduce the electronic structure of graphene and review general
aspects of magnetic moment formation. We then discuss various possibilities of experimentally
realizing magnetic moments coupled to graphene sheets, together with the relevant
microscopic descriptions.
Quite generically, this will lead to versions of the peudogap Kondo model, whose
theoretical treatment is discussed in some detail in Sec.~\ref{sec:pg}. We describe the
phase diagram, the quantum field theories and resulting critical properties, the
crossovers for finite chemical potential, and the implications for Kondo physics in
graphene.
Sec.~\ref{sec:exp} confronts these theoretical results with experimental data
obtained on impurity-doped graphene. Various real-world complications and their influence
on the interpretation of experiments will be discussed as well.
Finally, in Sec.~\ref{sec:out} we discuss quantum impurity problems which are relatives of
the graphene Kondo problem, thereby highlighting the generality of the theoretical
concepts developed in the field.


\section{Magnetic impurities in graphene}
\label{sec:imps}

In this section, we discuss how to realize magnetic impurity moments coupled to graphene
conduction electrons. To set the stage we first summarize basic aspects of both the
electronic structure of graphene and the formation of local moments in general.

\subsection{Electronic structure of graphene}

\begin{figure}[t!]
\centering
(a)\includegraphics[width=0.45\textwidth]{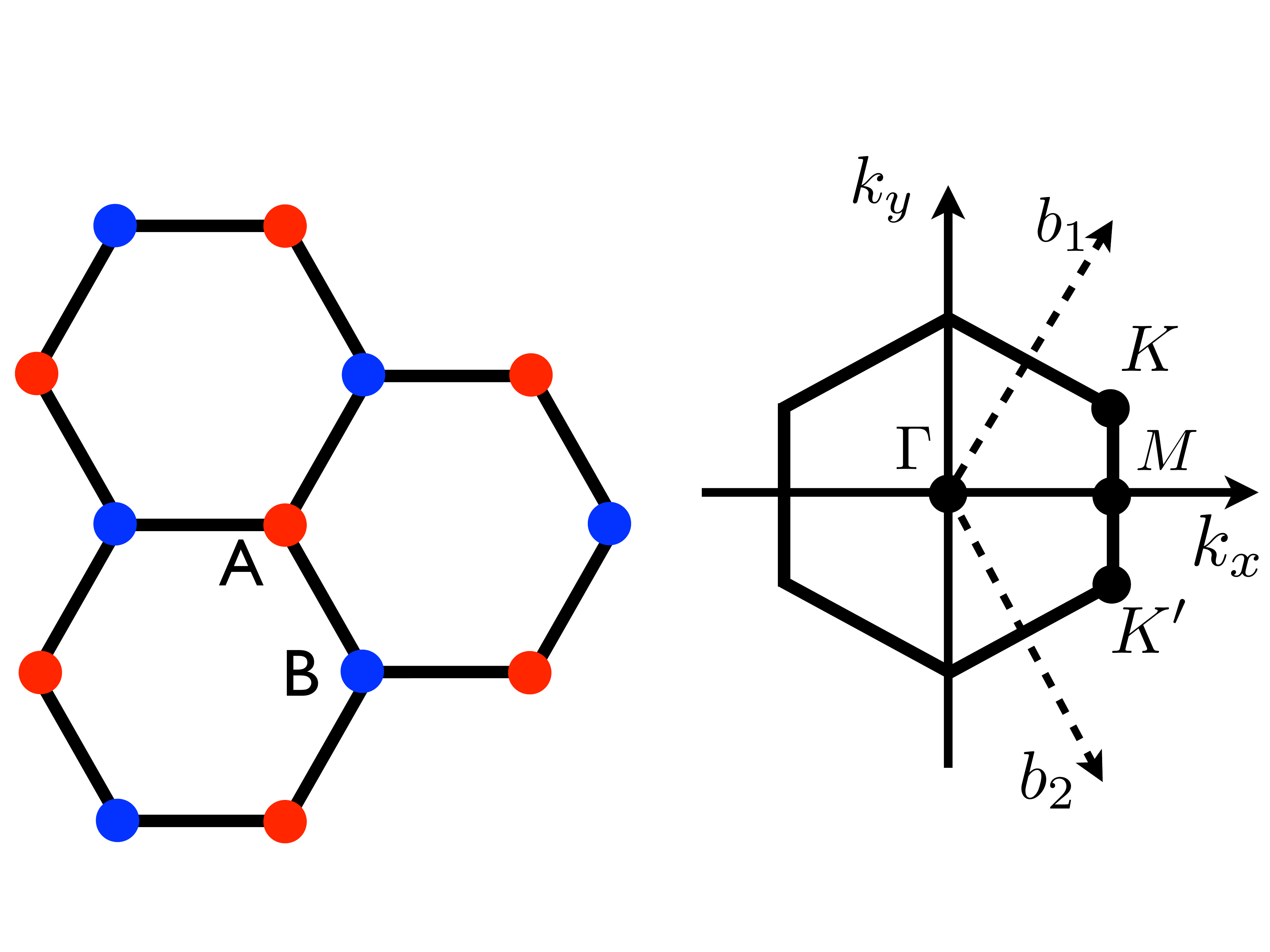}\hspace*{10mm}(b)\includegraphics[width=0.38\textwidth]{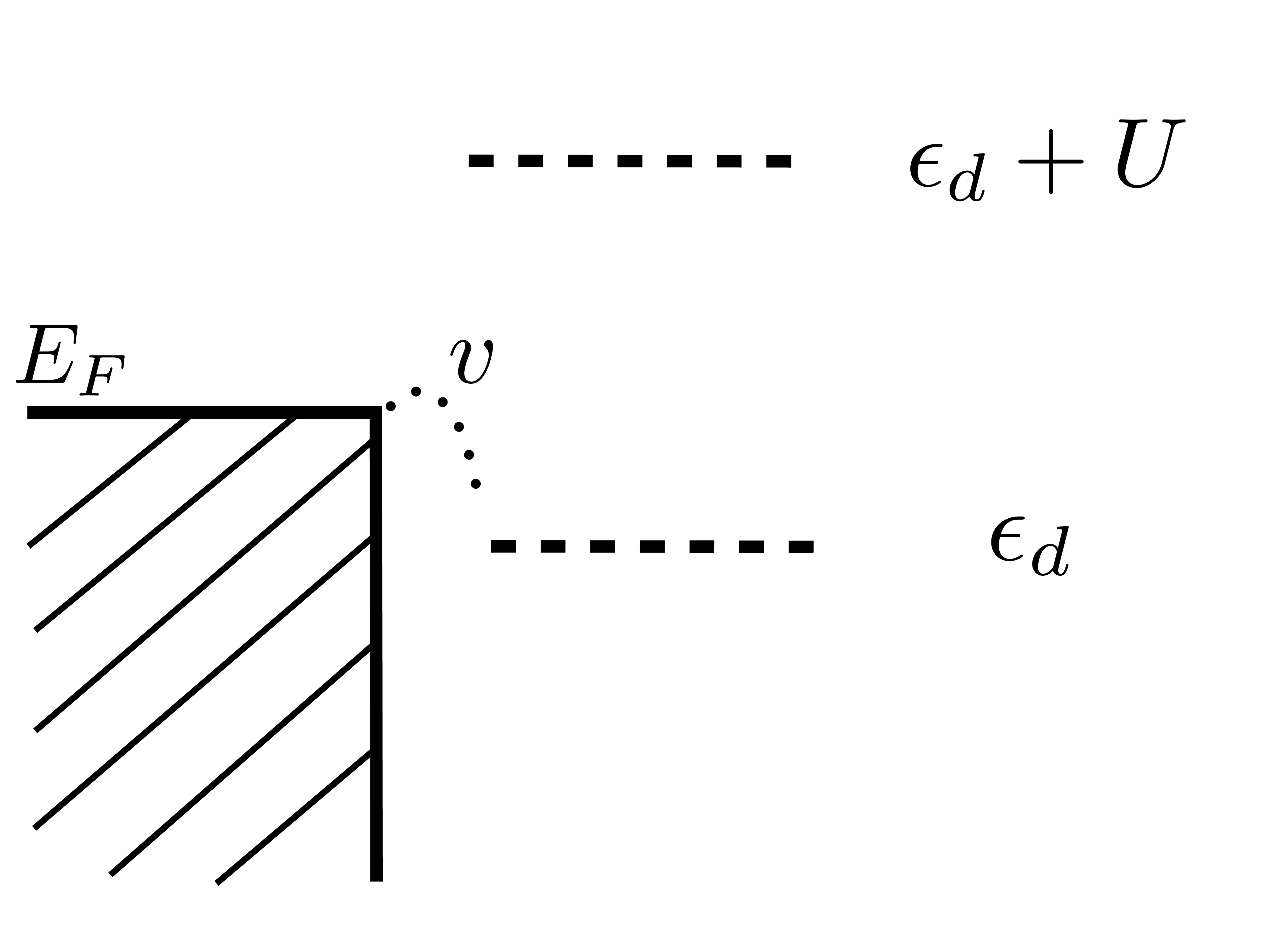}
\caption{
(a) Honeycomb lattice of graphene with two inequivalent
carbon atoms per unit cell, $A$ and $B$, and its hexagonal Brillouin zone.
The points $K$ and $K'$ are the touching points of the $E_\pm$
bands.
(b) Sketch of the AIM: A local spin-degenerate level with energy
$\epsilon_d$ is hybridized via $v$ with a sea of conduction electrons.
Local double occupancy costs the Coulomb energy $U$.
}
\label{fig:aim}
\end{figure}

Graphene is a 2d hexagonal arrangement of carbon
atoms~\cite{novo1,novo2,neto_rmp,sarma_rmp}. While the $sp^2$ orbitals hybridize to yield
the $\sigma$ orbitals which are electrically inert and responsible for the
remarkable mechanical robustness of graphene, its electronic structure is determined by
the $p_z$ orbitals which form the $\pi$-bonds. This
allows electron hopping between adjacent atoms and gives rise to a kinetic energy
described by the following tight-binding Hamiltonian~\cite{wallace47}:
\begin{equation}
\mathcal{H}_0 =-t\sum_{\langle i,j \rangle,\sigma}  \left( a^\dagger_{\sigma,i}
b^{\phantom{\dagger}}_{\sigma,j} +\textrm{h.c.} \right)-t' \sum_{\langle \langle i,j
\rangle \rangle,\sigma} \left( a^\dagger_{\sigma,i}
a^{\phantom{\dagger}}_{\sigma,j}+b^\dagger_{\sigma,i} b^{\phantom{\dagger}}_{\sigma,j}
+\textrm{h.c.} \right) \label{hhop}
\end{equation}
Here, $a_{\sigma,i}$ and $b_{\sigma,j}$ are annihilation operators for electrons on sites
$i,j$ of the two sublattices $A$ and $B$, see Fig.~\ref{fig:aim} a. The sums run over
pairs of nearest and next-nearest neighbours, respectively, with the hopping matrix
elements given by $t= 2.8$\,eV and $t' \approx 0.1t$.
Diagonalizing the Hamiltonian \eqref{hhop} yields two dispersive bands
\begin{eqnarray}
E_{\pm,\vec{k}} &=& \pm t \sqrt{3+f_{\vec{k}}}-t' f_{\vec{k}}~~~{\rm{with}} \; \nonumber \\
f_{\vec{k}}&=& 2\cos \left(\sqrt{3} k_y \al \right)+4 \cos
\left(\frac{\sqrt{3}}{2}k_y \al \right) \cos \left(\frac{3}{2}k_x \al \right)\;, \label{ek}
\end{eqnarray}
where $\al=1.42$\,\AA denotes the bond length, i.e. the distance between neighbouring carbon
atoms. The two bands (dubbed $\pi^*$ and $\pi$) touch at the two inequivalent
wavevectors $K= \left(\frac{2\pi}{3\al} , \frac{2 \pi}{3 \sqrt{3}\al} \right)$ and
$K'=\left(\frac{2\pi}{3\al} ,-\frac{2\pi}{3\sqrt{3}\al}  \right)$, see
Fig.~\ref{fig:aim}a.
Close to $K$ and $K'$ the dispersion is found to be linear:
\begin{equation}
E_{\pm,\vec{q}} = \pm v_F |\vec{q}|
\label{lindisp}
\end{equation}
where $\vec{q} = \vec{k} - \vec{K}$ (or $\vec{K}'$), $v_F=3t\al/2\approx
1.1\times 10^6$\,m/s, and an additional constant proportional to $t'$ has been
omitted in $E_\pm$.

For charge-neutral graphene, the electronic system is half-filled, i.e. in the ground
state all $E_-$ states are filled while the $E_+$ states are empty, such that the Fermi
level coincides with the band energy at $K$ and $K'$. Then Eq.~\eqref{lindisp} is the
dispersion of the system's low-energy excitations which admit a description in terms of
two-component massless Dirac equations, one for each of the valleys at $K$ and $K'$
\cite{neto_rmp}.
The 2d linear dispersion results in a low-energy density of states (DOS) per spin
which vanishes linearly at the Fermi level,
\begin{equation}\label{eq:dos}
\rho(\w) = \frac{2}{\sqrt{3}\pi t^2} |\w|\;,
\end{equation}
rendering graphene a semimetal. (Here and in the following, energies are measured
relative to the Fermi level, unless otherwise noted.)


\subsection{Local-moment formation in metals}
\label{sec:aim}

The general mechanism of local-moment formation in metals was formulated by Anderson in
1961~\cite{Anderson61}: a strong Coulomb interaction $U$ between electrons in a
spin-degenerate doublet of levels, $d_\sigma$, can freeze out charge fluctuations,
leaving behind an effective spin degree of freedom interacting with the spin density of
the conduction electrons, $c^{\phantom{\dagger}}_{{\vec{k}}\sigma}$.
Typically, such a situation is realized for impurity atoms with partially filled $d$ or
$f$ shells.
The corresponding minimal model is known as the Anderson impurity
model (AIM):
\begin{equation}\label{eq:AIM}
\mathcal{H} = \sum_{{\vec{k}},\sigma} \epsilon_{{\vec{k}}} c^\dagger_{{\vec{k}}\sigma}
c^{\phantom{\dagger}}_{{\vec{k}}\sigma} +\epsilon_d \sum_\sigma n_{d\sigma}
+ U n_{d\uparrow} n_{d\downarrow} + \sum_{{\vec{k}},\sigma}
\left( v_{\vec{k}} c^\dagger_{{\vec{k}}\sigma} d^{\phantom{\dagger}}_\sigma +\textrm{h.c.}
\right)
\end{equation}
where $n_{d\sigma} =  d^\dagger_\sigma d^{\phantom{\dagger}}_\sigma$.
Moment formation can be understood starting from the atomic limit, $v_{\vec{k}}=0$. For
$\epsilon_d<E_F$ and $\epsilon_d+U>E_F$ the $d$ level prefers single occupancy, such that
charge fluctuations are frozen out and an effective spin 1/2 degrees of freedom remains.
Upon switching on the hybridization $v_{\vec{k}}$, the so-formed local moment becomes
entangled with the conduction electrons.
It is convenient to convert the momentum dependence of
$v_{\vec{k}}$ and $\epsilon_{\vec{k}}$ into an energy-dependent hybridization function
\begin{equation}
\Delta(\w) = \sum_{\vec{k}} \frac{|v_{\vec{k}}|^2}{\w-\epsilon_{\vec{k}}}
\end{equation}
which fully characterizes the impurity's bath.
For small $v_{\vec{k}}$, one can utilize a Schrieffer-Wolff transformation to derive an
effective Kondo model from Eq.~\eqref{eq:AIM}, describing the interaction of the
local-moment spin with the conduction electrons, see Sec.~\ref{sec:pg}.


\subsection{Magnetic adatoms on graphene}
\label{sec:adatom}

We turn to the graphene-specific discussion of how to realize magnetic impurity
moments with sizeable electronic coupling to the host electrons.
An apparently straightforward route is to place a magnetic ad-atom onto the graphene
sheet, e.g., using the manipulation capabilities of a scanning tunneling microscope
(STM). For magnetic atoms like Fe or Co on the surface of conventional metals, this route
has been successfully used in the past to study local spectral signatures of Kondo
screening using STM techniques~\cite{eigler00}.

In the following we discuss theoretical aspects of such adatoms on graphene; experiments
will be reviewed in Sec.~\ref{sec:exp}.
The key questions for a quantitative understanding of the adatom's magnetism are:
\begin{enumerate}
\item{At which lattice position does the adatom adsorb? For graphene, possible
high-symmetry locations are shown in Fig.~\ref{fig:hyb}a and labeled h (hollow, in the center of a hexagon),
b (bridge, on a bond between two C atom), and t (on top of a C atom).}
\item{What is the spin state of the impurity adatom?}
\item{In an Anderson-model description, how are the impurity levels hybridized with conduction electrons?}
\end{enumerate}
All these question turn out vital for the presence or absence of the Kondo effect.
Answering these questions is highly non-trivial: While symmetries provide important
contraints on possible models \cite{wehling10,uchoa09,berakdar10,uchoa11,uchoa11b}, an
in-depth analysis requires ab-initio calculations, typically using variants of
density-functional theory (DFT).\footnote{ Notably, even for the ``classic'' situation of
Fe atoms in gold or silver, the correct model description has only been determined very
recently~\cite{Costi09} to be a spin-$\frac{3}{2}$ three-channel Kondo model. }

\subsubsection{Co.}
For Co atoms on an isolated graphene sheet, DFT calculations using the generalized
gradient approximation (GGA) found that the preferred adsorption is at site h with spin
$S=1/2$ \cite{wehling10}.
However, upon accounting for the local Coulomb repulsion $U$ and Hund's-rule coupling $J$
within the GGA+U method, this picture was modified: While for values of $U=2$\,eV and
$J=0.9$\,eV the h position and $S=1/2$ were still favoured, increasing $U$ to
4\,eV selected the t position and $S=3/2$~\cite{wehling10,wehling11}. As $U$ can
only be estimated to be in the range $2$\,eV$<U<4$\,eV, a clear-cut answer was missing
here, but the case of a spin $S=1/2$ on the h site was advertised as the most promising
candidate for observing Kondo physics.

In an alternative calculation, based on DFT augmented by a dynamical treatment of the 3d
levels in the framework of the one-crossing approximation, dubbed GGA+OCA, a
$S=3/2$ configuration on the t position was found to be most stable~\cite{kotliar10}. Here,
the authors argued in favour of a Kondo effect with full screening, as three
conduction-band channels coupled to the impurity spin.
Finally, in a refined quantum-chemical calculation~\cite{rudenko12} based on a complete
active-space self-consistent field approach, albeit on small clusters, it was found that
Co in an h position favours a higher-spin state of $S=3/2$.

Overall, the situation concerning Co adatoms is unclear at present, and more theory work
is called for. Specifically, the effect of the substrate, which possibly influences the
adatom's behaviour, has not been investigated so far.

\subsubsection{NiH.}
Recently, it has been proposed that a more promising route towards realizing the Kondo
effect could be provided by using NiH as adsorbing molecule~\cite{wehling11}. From
GGA+U, the molecule favors a $S=1/2$ state in the h position~\cite{wehling11}.

\subsubsection{Effective model.}
As in Sec.~\ref{sec:aim}, the physics of a localized impurity level hybridized with
graphene electrons can be described by an Anderson impurity model,
\footnote{The mean-field solution of the Anderson model for charge-neutral graphene has
been discussed in Ref.~\cite{uchoa08}; for aspects of the full solution see
Ref.~\cite{GBI} and Sec.~\ref{sec:pg}.}
which then may be mapped onto an effective Kondo model, see Sec.~\ref{sec:pg}.

\begin{figure}
\centering
(a)\includegraphics[width=0.30\textwidth]{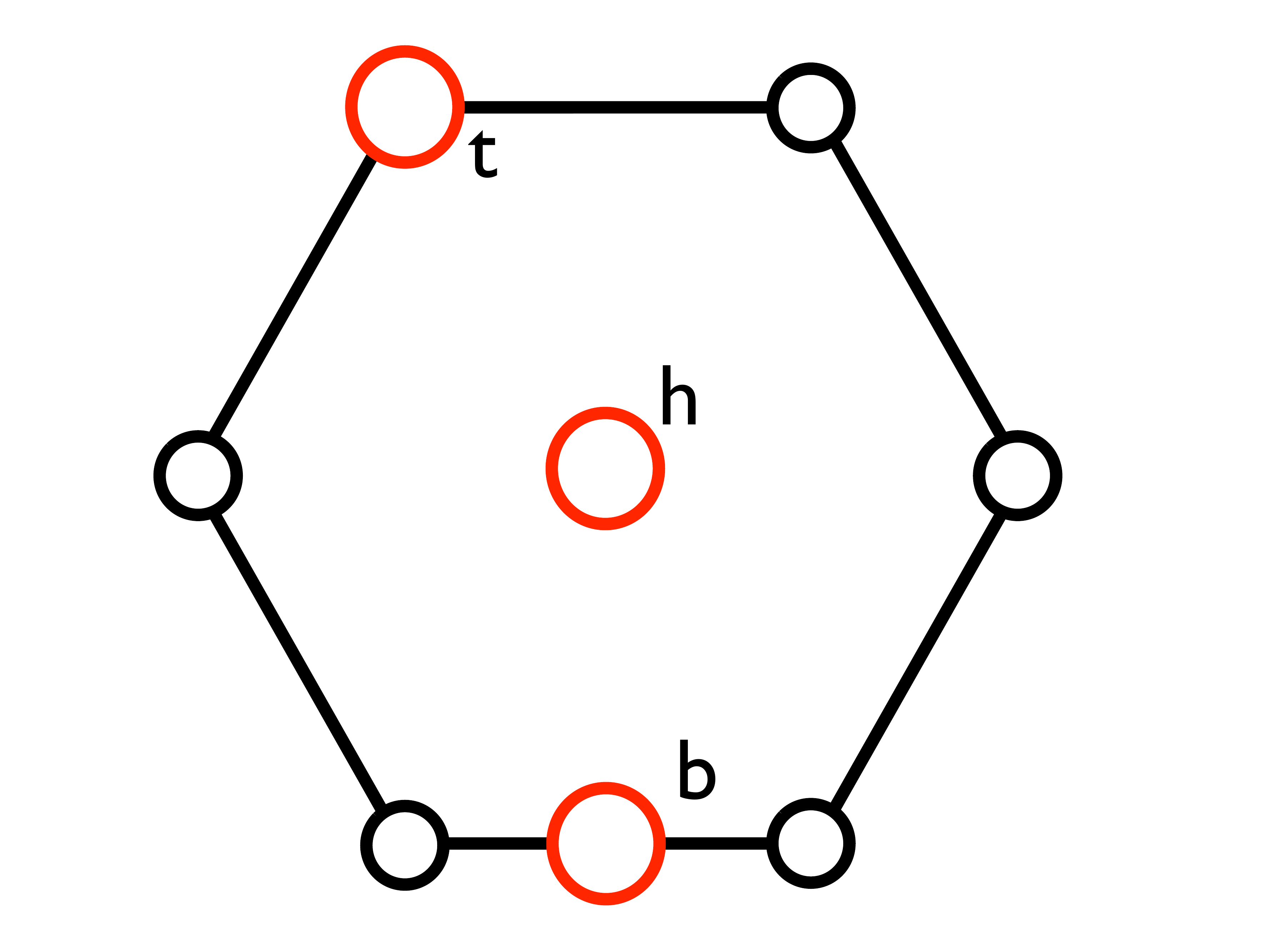}(b)\includegraphics[width=0.58\textwidth]{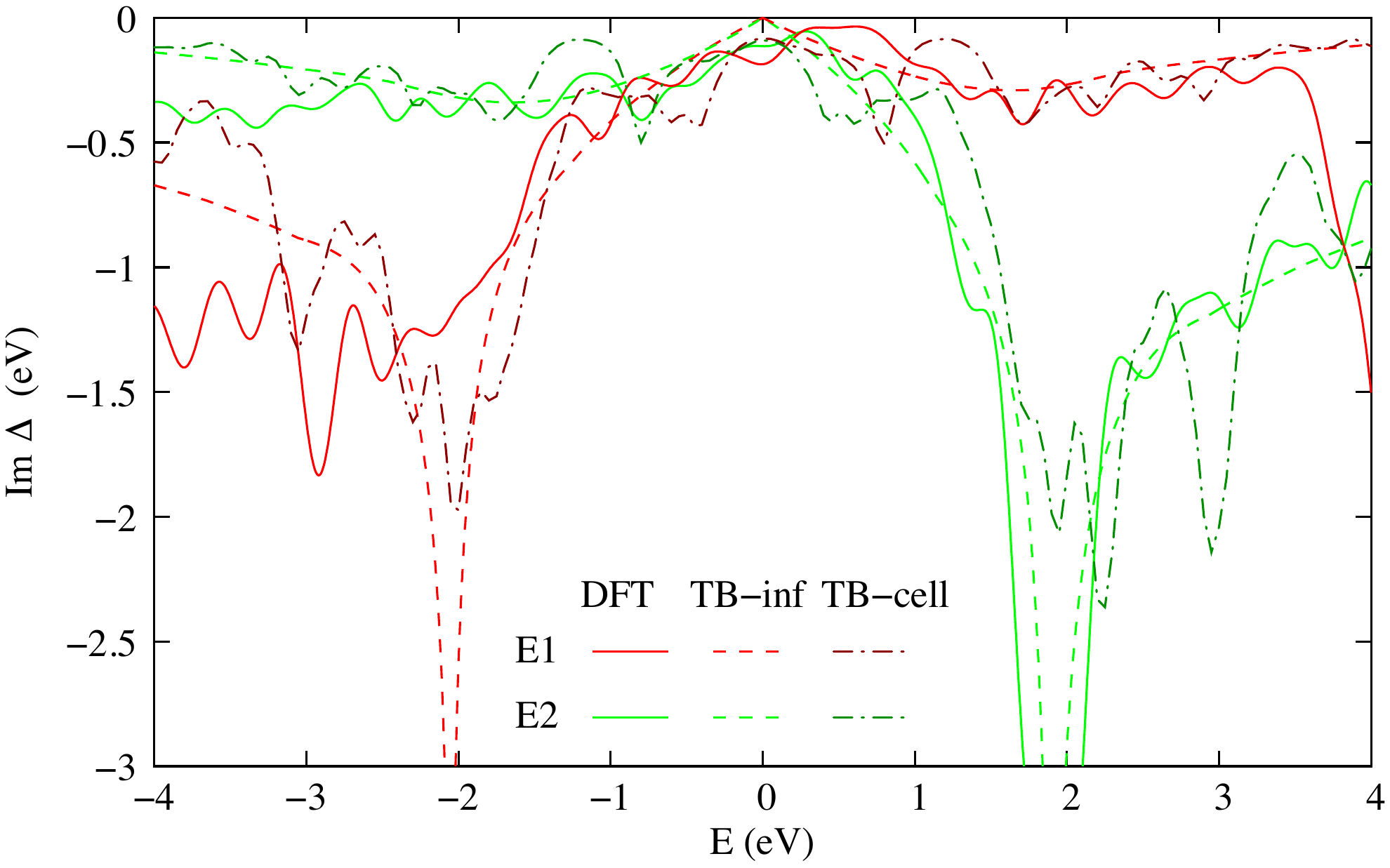}
\caption{
(a) Possible high-symmetry adsorption sites of ad-atoms on the graphene sheet, denoted by
h (hollow), t (top), and b (bridge).
(b) Hybridization function of the active impurity level of a Co atom in the h position
in a spin state of $S=1/2$ \cite{wehling10}. Solid: DFT result, dashed: tight-binding fit.
E1, E2 refer to different orbital configurations, with E1 being lower in energy
(taken from Ref.~\cite{wehling10}).

}
\label{fig:hyb}
\end{figure}

For a spin-$\frac{1}{2}$ Co atom in the h position, it was argued \cite{wehling10} that,
due to an approximate orbital degeneracy, the impurity behaviour at elevated energies
corresponds to that of an SU(4) Kondo effect. The orbital splitting is roughly $60$\,meV, such
that a standard single-channel SU(2) $S=1/2$ Kondo or Anderson model applies at energies
below this scale, albeit with a non-standard hybridization function $\Delta(\w)$. The
latter  was calculated in Ref.~\cite{wehling10} and is reproduced in Fig.~\ref{fig:hyb}b. It
vanishes near the Dirac-point energy (set to zero here) according to
\begin{eqnarray}
{\rm Im}\,\Delta(\omega) \propto |\omega|,
\label{pghyb}
\end{eqnarray}
a behaviour inherited from the graphene DOS, while for higher energies there is sizeable
particle--hole asymmetry. Both features turn out vital for the Kondo effect, see
Sec.~\ref{sec:pd}.
It can be expected on symmetry grounds that this effective model with a similar hybridization function
applies to NiH on the h site as well.

\subsection{Defect-induced moments in graphene}
\label{sec:vac}

A different route towards magnetic impurities in graphene is via point defects which
themselves induce moments. Relevant defects include vacancies, created e.g. by
irradiation~\cite{grigorieva12,kawakami12}, and hydrogen and fluorine adatoms.

\subsubsection{$\pi$-orbital magnetic moment.}

Removing a single site from the $\pi$-electron tight-binding Hamiltonian in
Eq.~\eqref{hhop} induces a single localized state {\em at} the Dirac-point energy for
$t'=0$, which becomes a quasi-localized resonance near the Dirac-point energy for
non-zero $t'$~\cite{neto_rmp,pereira06}. Consequently, it has been proposed that, upon
including Coulomb interactions, a magnetic moment may be formed in this localized state,
i.e. in the vicinity of the vacancy. However, in real graphene the lattice near the
vacancy will reconstruct, such that determining the proper effective model requires
ab-initio studies.

Notably, the results of those studies are again controversial. Initially, local-moment
formation for a vacancy was confirmed using DFT, however, the coupling to the conduction
electrons was argued to be ferromagnetic~\cite{helm06}, such that no Kondo effect can be
expected. A similar conclusion was drawn from a study using dynamical mean-field
theory (DMFT)~\cite{pruschke11} where a Curie-type susceptibility was found indicative of
a free moment.
In contrast, for an H atom adsorbed on top of a C atom, Ref.~\cite{balseiro12} argued
that the physics can be described by a single $S=1/2$ impurity coupled
antiferromagnetically to the environment, such that screening with sizable Kondo temperature
should be possible.
Finally, a more recent DFT calculation~\cite{yndurain12} for a finite concentration of H
adatoms concluded that, while there can be localized $\sigma$-orbital moments,
$\pi$-orbital moments only occur for the unlikely situation of hydrogenation of all
dangling $\sigma$ bonds. This paper pointed out the importance of considering both $\sigma$
and $\pi$ bonds and their reconstruction around a vacancy.

In any case, if a $\pi$ moment forms then it can be expected that its coupling to the
conduction electrons is described by an Anderson/Kondo model with a pseudogap hybridization
function as in Eq.~\eqref{pghyb}, possibly with a large particle--hole asymmetry at
higher energies due to potential scattering. The physics of this pseudogap Kondo model will be
discussed in detail in Sec.~\ref{sec:pg}.

\subsubsection{$\sigma$-orbital magnetic moment.}

An interesting alternative is to consider the carbon's $\sigma$ orbitals. While
local-moment formation driven by Coulomb interaction is possible here as well, the
obstacle is that -- for flat graphene -- the hybridization between $\sigma$ and $\pi$
orbitals vanishes, i.e. such a moment would not couple to the conduction electrons.
However, hybridization of $\sigma$ and $\pi$ orbitals becomes possible once structural
deviations from the flat geometry are included, i.e. by corrugations of the
graphene sheet around the impurity site.

Ref.~\cite{cazalilla12} investigated vacancies and $\sigma$-orbital moments from an LDA+U
perspective. It was found that $S=1/2$ or $S=1$ $\sigma$ moments emerge, which can have
sizeable single-channel hybridization to the $\pi$ electrons upon including rippling,
which occurs under small isotropic compression around reconstructed vacancies.
Interestingly, the hybridization function within an effective Anderson model was found
to be low-energy divergent according to
\begin{eqnarray}
{\rm Im}\,\Delta(\omega) \propto \frac{1}{|\omega|\ln^2 |\omega/D|}\,,
\end{eqnarray}
where $D$ is the bandwidth, providing a high-energy cutoff. In that situation the Kondo
temperature can largely be enhanced due to the massive density of states at low energies
\cite{cazalilla12,VB02}, but detailed studies of this model are not available.


\section{The pseudogap Kondo problem}
\label{sec:pg}

We now discuss the rich physics of the so-called pseudogap Kondo model, relevant to
low-energy behaviour of magnetic moments in graphene. We will restrict our
attention to the case of a spin $S=1/2$ coupled to a single screening channel; the
two-channel version will be briefly mentioned in Sec.~\ref{sec:pg2ck}.

\subsection{The pseudogap Kondo model}

The standard Kondo Hamiltonian~\cite{hewson} reads
\begin{eqnarray}\label{eq:Kondomodel}
\mathcal{H} = \sum_{{\vec{k}},\sigma} \epsilon_{{\vec{k}}} c^\dagger_{\vec{k}\sigma}
c^{\phantom{\dagger}}_{\vec{k}\sigma} +V_0\sum_{{\vec{k}},{\vec{k}}',\sigma}
c^\dagger_{\vec{k}\sigma} c^{\phantom{\dagger}}_{\vec{k}'\sigma}  +J_0\; {\vec{S}}
\cdot {\vec{s}}_0,
\end{eqnarray}
where the notation follows Sec.~\ref{sec:aim}, $\vec{S}$ is the impurity spin
$S=1/2$, and ${\vec{s}}_0=\frac{1}{2}\sum_{{\vec{k}}{\vec{k}}'} c^{\dagger}_{\vec{k}\sigma}
{\vec{\tau}}_{\sigma \sigma'}c^{\phantom{\dagger}}_{\vec{k}'\sigma'}$ is the
conduction-electron spin density at the impurity site, with ${\vec{\tau}}$ the vector of
Pauli matrices. The Kondo coupling $J_0$ and the potential-scattering strength $V_0$
characterize the impurity. If the Kondo model is derived from the more general Anderson
model, Eq.~\eqref{eq:AIM}, in the limit of small charge fluctuations, second-order
perturbation theory yields \cite{hewson}:
\begin{equation}
J_0 = 2 v^2 \bigg(\frac{1}{|\epsilon_d|} + \frac{1}{|U+\epsilon_d|}\bigg),~~
V_0 = \frac{v^2}{2} \bigg(\frac{1}{|\epsilon_d|} - \frac{1}{|U+\epsilon_d|}\bigg)
\label{swo}
\end{equation}
where $v_{\vec{k}}\equiv v$ has been assumed.
Note that $V_0\neq 0$ breaks particle--hole symmetry if the host DOS is particle--hole
symmetric, $\rho(\w) = \rho(-\w)$.\footnote{This symmetry is not obeyed even for neutral
graphene due to finite next-neighbor hopping $t'$.}

For a metallic host, the DOS $\rho(\w)$ is finite at the Fermi level. Then, for
antiferromagnetic $J>0$ the impurity spin is screened below the so-called Kondo
temperature $\TK$. For a flat conduction-band DOS, $\rho(\w)=\rho_0$, one finds~\cite{hewson}:
\begin{eqnarray}
\label{tkeq}
\TK= \sqrt{D J_0}\, e^{-1/(J_0 \rho_0)}\;.
\end{eqnarray}
Importantly, in this metallic Kondo problem, the crossovers at finite energies and
temperatures are characterized by the {\em single} scale $\TK$. For instance,
the impurity susceptibility displays single-parameter scaling: $\chi_{\rm
imp}(T)$ is a universal function of $T/\TK$ only, and does not depend on further microscopic
details.

In the following, we will instead concentrate on the case of a pseudogap DOS,
\begin{equation}
\rho(\omega) = \frac{1+r}{2D^{r+1}}\,|\omega|^r\,\Theta(|\w|-D)
\label{pgdos}
\end{equation}
with $r>0$. This is the situation of a semimetal with vanishing
DOS at the Fermi level. Consequently, the tendency toward Kondo screening is reduced,
such that no screening occurs at small Kondo coupling $J_0$. As a result,
a quantum phase transition between phases without and with screening occurs upon
increasing $J_0$ \cite{mv_rev,withoff}, as discussed in detail below.
Importantly, the form of the DOS \eqref{pgdos} with $r=1$ is relevant for both $d$-wave
superconductors and charge-neutral graphene at low energies.
We note that the implications of the vanishing DOS for the x-ray edge problem in graphene
and the associated Anderson orthogonality catastrophe were discussed in
Ref.~\cite{hentschel07}.

\subsection{Phase diagram}
\label{sec:pd}

Despite its simplicity, the pseudogap Kondo model has an extraordinarily rich phase
diagram, first determined by Gonzalez-Buxton and Ingersent \cite{GBI} using Wilson's
numerical renormalization group (NRG) technique. The physics depends not only on $J_0$
and the exponent $r$ of the low-energy DOS, but also on the presence or absence of
particle--hole symmetry.
[We recall that particle--hole symmetry requires {\em both} $\rho(\w) =
\rho(-\w)$ in the host and $U=-2\epsilon_d$ in the AIM \eqref{eq:AIM}.]

Thanks to both numerical \cite{GBI,bulla97,bulla00,si02} and perturbative \cite{VF04,FV04} RG
studies, the phases and phase transitions of the pseudogap Kondo and Anderson models are
by now fully understood, and will be summarized below. This discussion, restricted to
$r> 0$, mainly follows Ref.~\cite{GBI} and is cast in the language of RG flows and
fixed points. We start by listing the fixed points of the problem. For brevity, we will
use the acronyms of Ref.~\cite{GBI} for both the phases and their stable fixed points.

\paragraph{Local-moment phase \bf{LM}.}
Here the impurity moment is asymptotically decoupled from the host and behaves like a
free local moment, i.e. it has a residual entropy $\Simp=\ln2$.
\paragraph{Symmetric strong-coupling phase \bf{SSC}.}
This phase corresponds to Kondo screening in the presence of particle--hole symmetry and
is the generalization of the metallic Kondo-screened phase to finite $r$.
Interestingly, SSC is characterized by a residual entropy $\Simp=2r\ln
2$, which means the impurity moment is only partially screened.
\paragraph{Asymmetric strong-coupling phase \bf{ASC}.}
In the absence of particle--hole symmetry and for $r>0$, full screening with $\Simp=0$
obtains at ASC, which is maximally particle--hole asymmetric.
\paragraph{Critical points {\bf{SCR}} and {\bf{ACR}}.}
The pseudogap Kondo problem has two distinct critical fixed points, distinguished by
their symmetry under particle--hole transformation and denoted ``symmetric critical''
(SCR) and ``asymmetric critical'' (ACR), respectively. They both exist over a restricted
range of $r$ values.
\bigskip

\begin{figure}[t]
\includegraphics[width=0.33\textwidth]{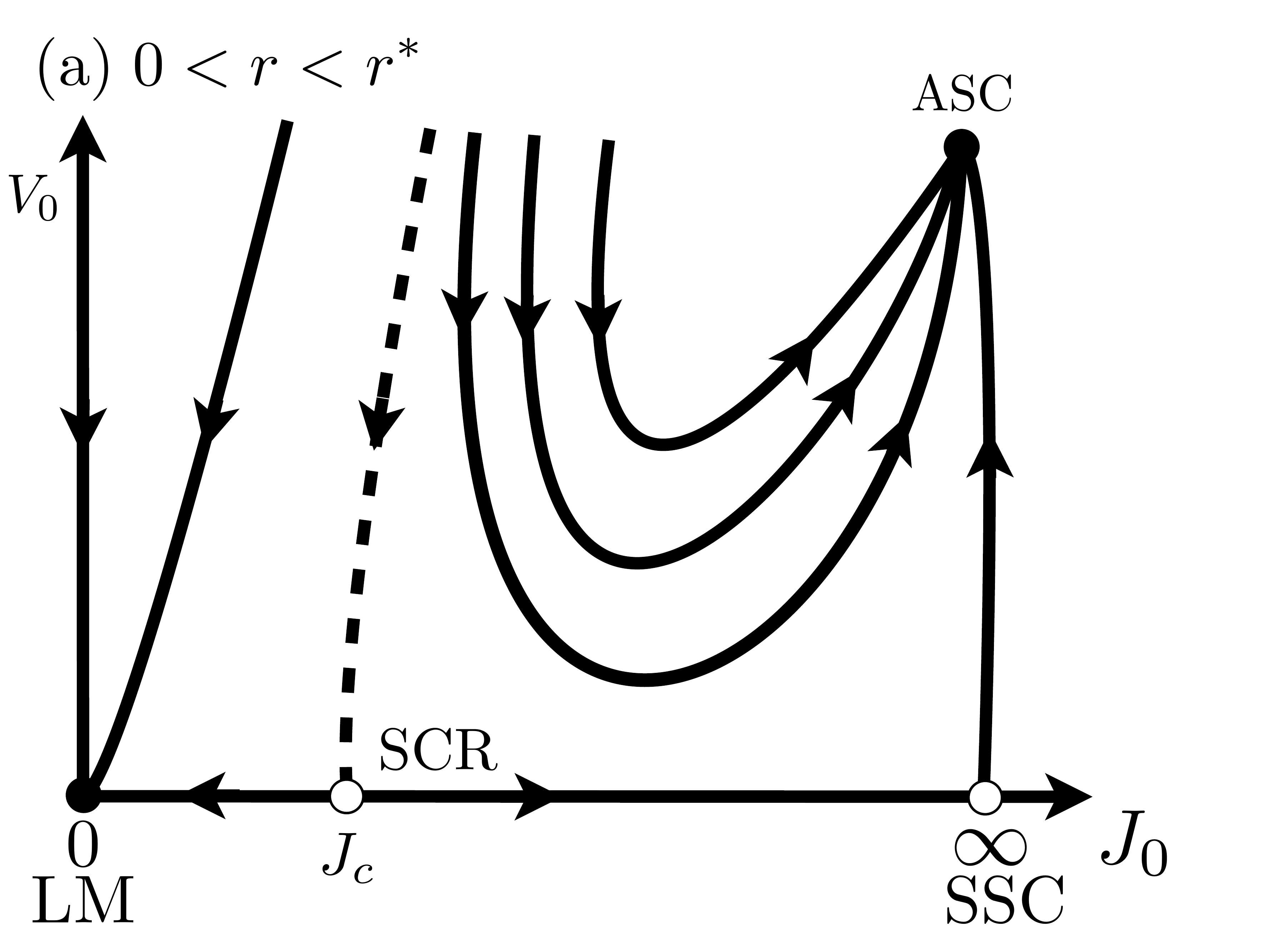}
\includegraphics[width=0.33\textwidth]{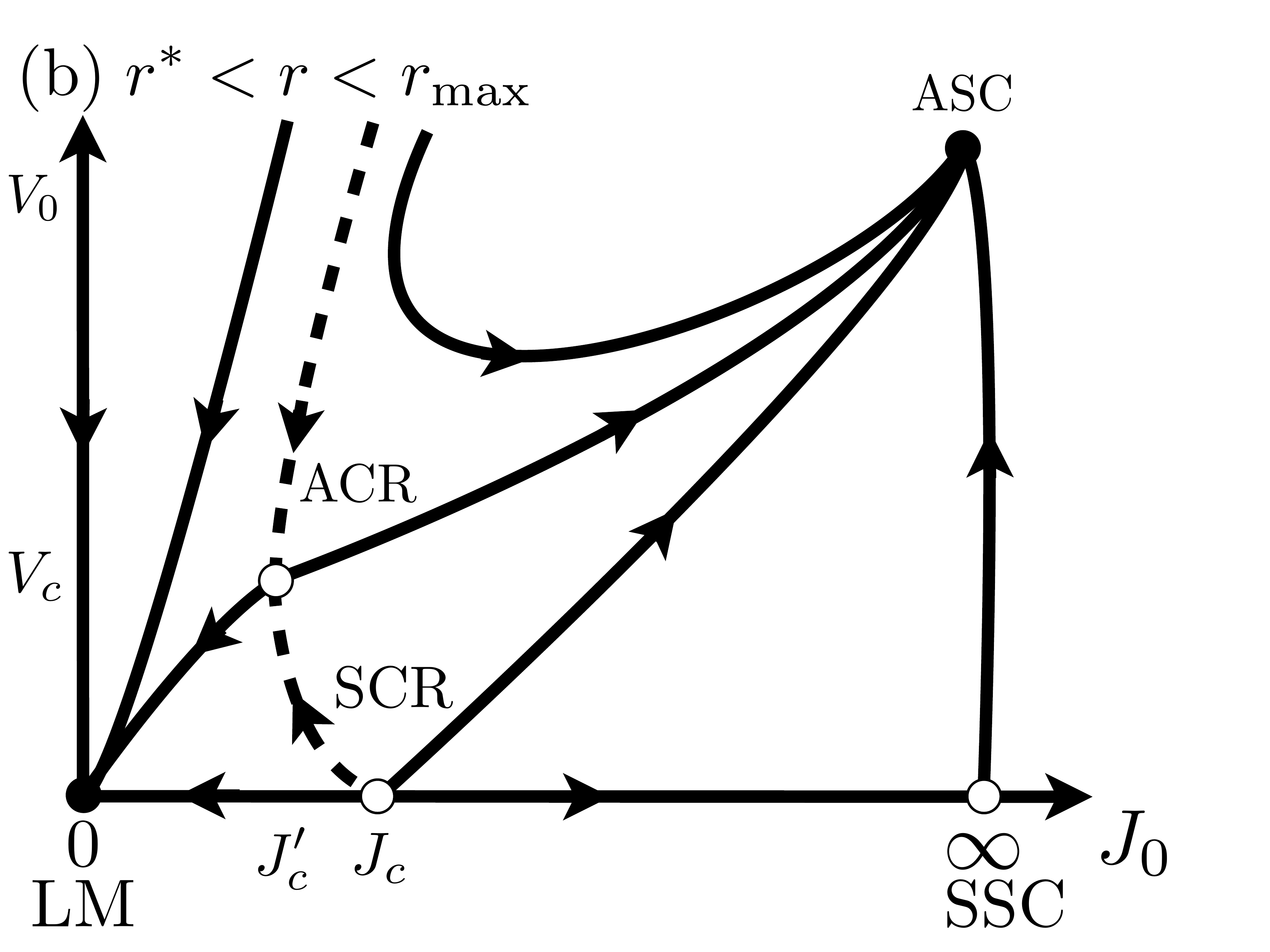}
\includegraphics[width=0.33\textwidth]{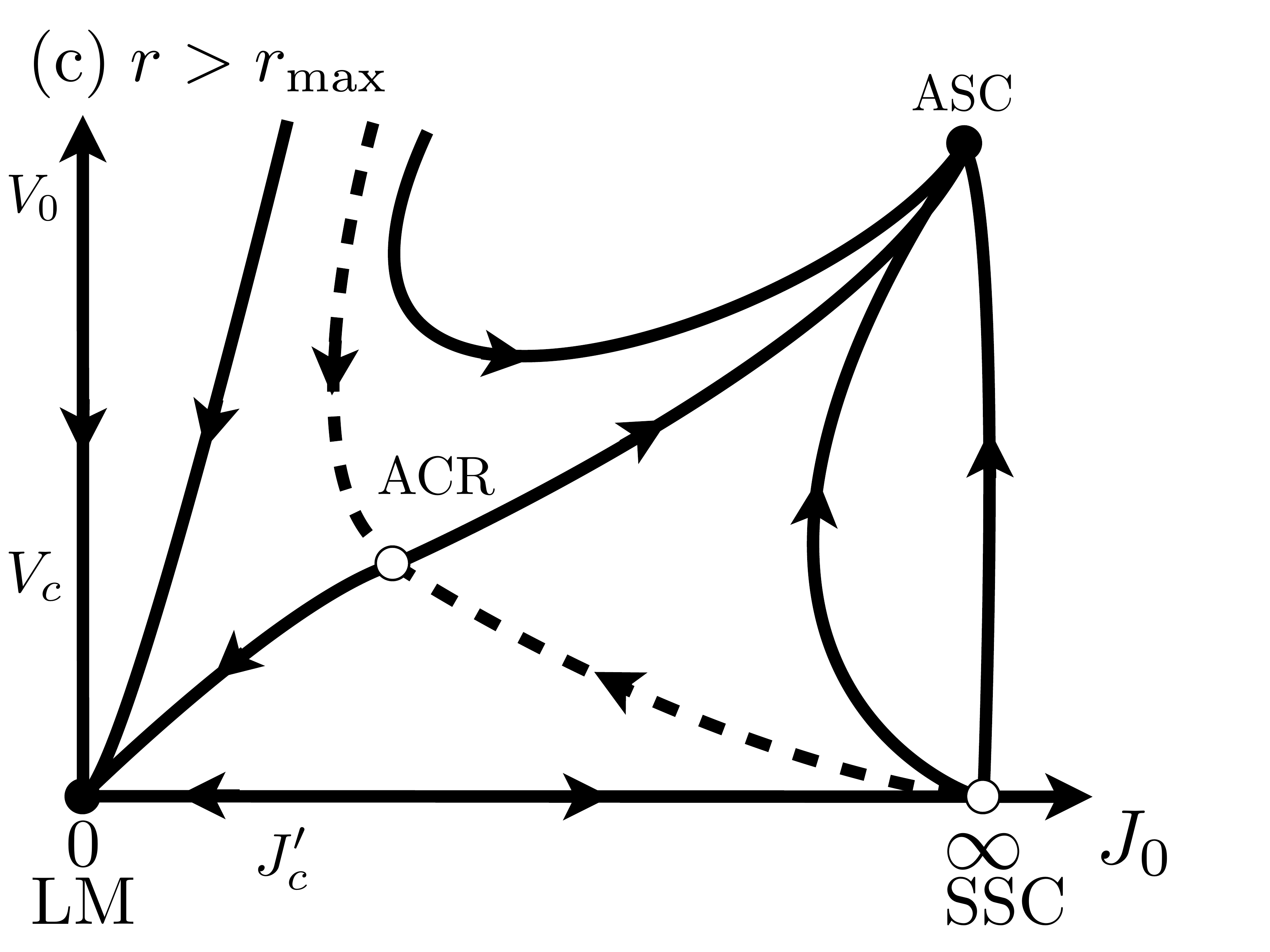}
\caption{
Schematic RG flow diagrams for the pseudogap Kondo model \cite{GBI} in the plane spanned by the Kondo
coupling $J_0$ and the potential scattering $V_0$, the latter measuring particle--hole asymmetry.
The flow topology changes qualitatively as function of $r$, as shown in the three panels,
with $r^\ast = 0.375\pm0.002$ and $\rmax=1/2$, for details see text.
Full dots denote stable fixed points, while open dots are critical fixed points. Dashed lines
denote separatrices, i.e., phase boundaries.
}
\label{gonzalezflow}
\end{figure}

As deduced from the numerical solution of the pseudogap Kondo model \cite{GBI}, the topology of the
phase diagram changes {\em qualitatively} as the bath exponent $r$ is varied.
Different phase diagram topologies are observed in three regimes, see Fig.~\ref{gonzalezflow}.

\paragraph{a) $\bf{0<r<r^* = 0.375\pm0.002}$}
\begin{enumerate}
\item{For particle--hole symmetry, a critical coupling $\Jc$, associated with SCR, separates LM from SSC. For initial values $J<\Jc$ the flow is directed towards LM, whereas for $J>\Jc$ the flow is directed towards SSC.}
\item{For finite particle--hole asymmetry, i.e. $V_0\neq 0$, there is a separatrix which separates the flow towards LM from the flow towards ASC.}
\item{Particle-hole asymmetry is irrelevant at LM and SCR while it is relevant at SSC where it drives the flow towards ASC. SCR is thus a multicritical fixed point.}
\end{enumerate}
\paragraph{b) $\bf{r^*<r<\rmax=1/2}$}%
\begin{enumerate}
\item{For $V_0=0$ there still exists a critical coupling which separates LM from SSC.}
\item{SCR is now unstable w.r.t. particle--hole asymmetry, and a new asymmetric critical fixed point ACR emerges, controlling the transition between LM and ASC.}
\end{enumerate}
\paragraph{c) $\bf{r>\rmax}$}
\begin{enumerate}
\item{SCR merges with SSC, such that there is no Kondo screening at particle--hole symmetry, irrespective of the strength of the Kondo coupling $J_0$.}
\item{Screening is possible for finite asymmetry, where ACR continues to control the LM--ASC transition. }
\item{The critical exponents at ACR take trivial values for $r>1$, such that $r=1$ -- the case relevant for charge-neutral graphene -- acts as an upper-critical dimension \cite{GBI,si02,VF04}.}
\end{enumerate}
\paragraph{d) $\bf{-1<r<0}$}

This regime can possibly be realized in the case of reconstructed vacancies in
graphene~\cite{cazalilla12} but was analyzed more generally in Ref.~\cite{VB02}. SSC is
stable, and a critical point ACR separates SSC from a newly emerging fixed point ALM,
located at $J_0=0$ and $V_0=\infty$. In the following we will, however, not discuss $r<0$
in any detail.

\smallskip

Notably, the pseudogap Kondo and Anderson models share identical fixed points and quantum
phase transitions \cite{GBI}. This observation can be rationalized within the effective
field theories described in Sec.~\ref{sec:field} below.


\subsection{Slave-boson mean-field theory}

A simple and popular approach to the Kondo model in Eq.~\eqref{eq:Kondomodel} is the
slave-boson mean-field theory \cite{hewson,readnewns}; a very similar mean-field theory
can be applied to the Anderson model \eqref{eq:AIM}. In this approach, formally justified
in a limit where the spin symmetry is taken to be SU($N$) with $N\to\infty$, the Kondo
interaction is decoupled by a Hubbard-Stratonovich field which is then approximated to be
static. This results in a renormalized free-particle Hamiltonian which reproduces salient
low-temperature properties of a Kondo-screened impurity in a metal.

The slave-boson approach has been applied to the pseudogap Kondo model in numerous of
papers \cite{uchoa11b,uchoa08,withoff,si98,cassa,tolya01,tolya02,zhuang09,dellanna10}. It
reproduces the existence of a quantum phase transition at $r>0$, however, the critical
properties of this transition are only correctly captured for small $r$.
For larger $r$ including the graphene case $r=1$, the slave-boson method becomes
unreliable. It is not sensitive to the subtle effects of particle--hole symmetry
breaking: It fails to describe the properties near the ACR fixed point, and it
incorrectly predicts a phase transition for $r>1/2$ even in the particle--hole symmetric
case.
Therefore, quantitative calculations require numerical methods such as NRG.


\subsection{Critical field theories}
\label{sec:field}

The complicated topology of the RG flow, Fig.~\ref{gonzalezflow}, suggests that different
field theories are required to describe the critical properties near the SCR and ACR
fixed points. Such field theories have been worked out in detail in
Refs.~\cite{VF04,FV04} and provide an essentially complete analytical understanding of
the pseudogap Anderson and Kondo models. Interestingly, none of these field theories is
of conventional (i.e. bosonic) Landau-Ginzburg-Wilson type; instead all are of genuinely
fermionic character and are formulated in the degrees of freedom of either the Kondo or the
Anderson model.

In the following we shall summarize the three relevant critical theories. As will become
clear, only the third will be appropriate to describe the quantum phase transition of
Kondo impurities in charge-neutral graphene, where $r=1$.
When specifying flow equations from perturbative RG, we will assume a symmetric pseudogap
density of states as in Eq.~\eqref{pgdos}. The effect of a high-energy particle--hole
asymmetry in the DOS can absorbed in the impurity part of the Hamiltonian, e.g., the
potential scattering term of the Kondo model. This can be rationalized within RG, where
integrating out the particle--hole asymmetric piece of the bath at high energies yields
an effective model with particle-hole symmetric bath at low energies and a renormalized
impurity Hamiltonian, where in particular the particle--hole asymmetry is accumulated.

\subsubsection{SCR: Kondo model.}
For small $r$, an efficient description of the physics at SCR is obtained via the
Kondo model itself, Eq.~\eqref{eq:Kondomodel}. A perturbative expansion can be performed
in $J_0$ and $V_0$ around the LM fixed point where $J_0=V_0=0$ \cite{withoff,si98,si02,FV04}.
As is standard practice, we
introduce dimensionless couplings $j$ and $v$, for details see Ref.~\cite{FV04}. Power
counting reveals that both couplings are marginal for $r=0$ and irrelevant for $r>0$,
$\dim[j] = \dim[v] = -r$. The one-loop flow equations read
\begin{eqnarray}
\frac{dj}{d \ln D}&=&r j -j^2 \quad {\rm{and}} \quad \frac{dv}{d \ln D}=r v \;,
\label{poorflow}
\end{eqnarray}
where $D$ denotes the running UV cutoff, initially set by the width of the host band.
Eq.~\eqref{poorflow} yields a critical fixed point (SCR) at $j^\ast=r+\mathcal{O}(r^2)$,
$v^\ast=0$, which separates the flows towards weak and strong coupling. Controlled
calculations near SCR are therefore possible in a double expansion in $r$ and $j$.
Potential scattering is irrelevant at SCR and consequently does not play a role for
leading critical exponents.

Comparing these properties with the numerically deduced flows in Fig.~\ref{gonzalezflow},
it is clear that this Kondo description of SCR is appropriate for $0<r<r^*$. It does,
however, not capture the physics for $r>r^\ast$ where $v$ becomes a relevant perturbation
at the SCR fixed point, and it is obviously inappropriate for the graphene case $r=1$.

A perturbative calculation of static critical properties of SCR using the Kondo expansion
indeed shows excellent agreement with NRG results for small $r$~\cite{FV04}. Crossover
functions and dynamical properties have been studied as well, using a combination of
perturbative RG and Callan-Szymanzik equations \cite{FFV06}, and agreement with NRG
results has been found where those are available.\footnote{In the quantum-relaxational
finite-temperature regime of $\w\ll T$ numerical studies are notoriously difficult. Here,
the Callan-Szymanzik results \cite{FFV06} disagree with numerical data~\cite{glossop11}:
the latter indicate that the imaginary part of local Green function (or T matrix)
$G''(\omega,T)$ multiplied by $T^r$ goes to a non-zero constant, i.e. $T^r G''(\omega/T
\to 0)=c\neq0$, while the former suggests that it goes to zero.}

\subsubsection{SCR: Symmetric Anderson model.}

The Anderson impurity model, originally introduced as model for local-moment formation
(Sec.~\ref{sec:aim}), turns out to provide the relevant degrees of freedom to describe
pseudogap Kondo criticality for all $r>0$ \cite{FV04}. This generically implies that critical
fluctuations occur not only in the spin channel, but also in the charge channel
\cite{pixley12}.

To discuss the critical behaviour near SCR, we consider a symmetric AIM,
Eq.~\eqref{eq:AIM} with $\epsilon_d = -U/2$, a momentun-independent hybridization $v$,
and a particle--hole symmetric bath DOS as in Eq.~\eqref{pgdos}. The point
$\epsilon_d=U=v=0$ is referred to as the free-impurity fixed point (FImp), whereas the
parameter sets $v=0$ and $\epsilon_d = -U/2= \pm\infty$ correspond to doubly degenerate
local-moment states in the charge and spin channel, respectively. Therefore $v=0$,
$\epsilon_d=-\infty$ can be identified with the LM fixed point, while $v=0$,
$\epsilon_d=\infty$ is dubbed LM'.

Notably, the Anderson model is exactly solvable for any $v$ at $U = 0$, known as
resonant-level model. In the particle--hole symmetric case, its low-energy physics can be
identified with that of the SSC fixed point introduced above:\footnote{In the metallic
case, $r=0$, it is known that the resonant-level model is the correct fixed-point theory
for a Kondo-screened impurity spin.} its properties correspond to a partial screening of
the impurity degrees of freedom, with a residual entropy of $\Simp = 2r \ln 2$
\cite{GBI,FV04}.

\begin{figure}[t]
\includegraphics[width=0.33\textwidth]{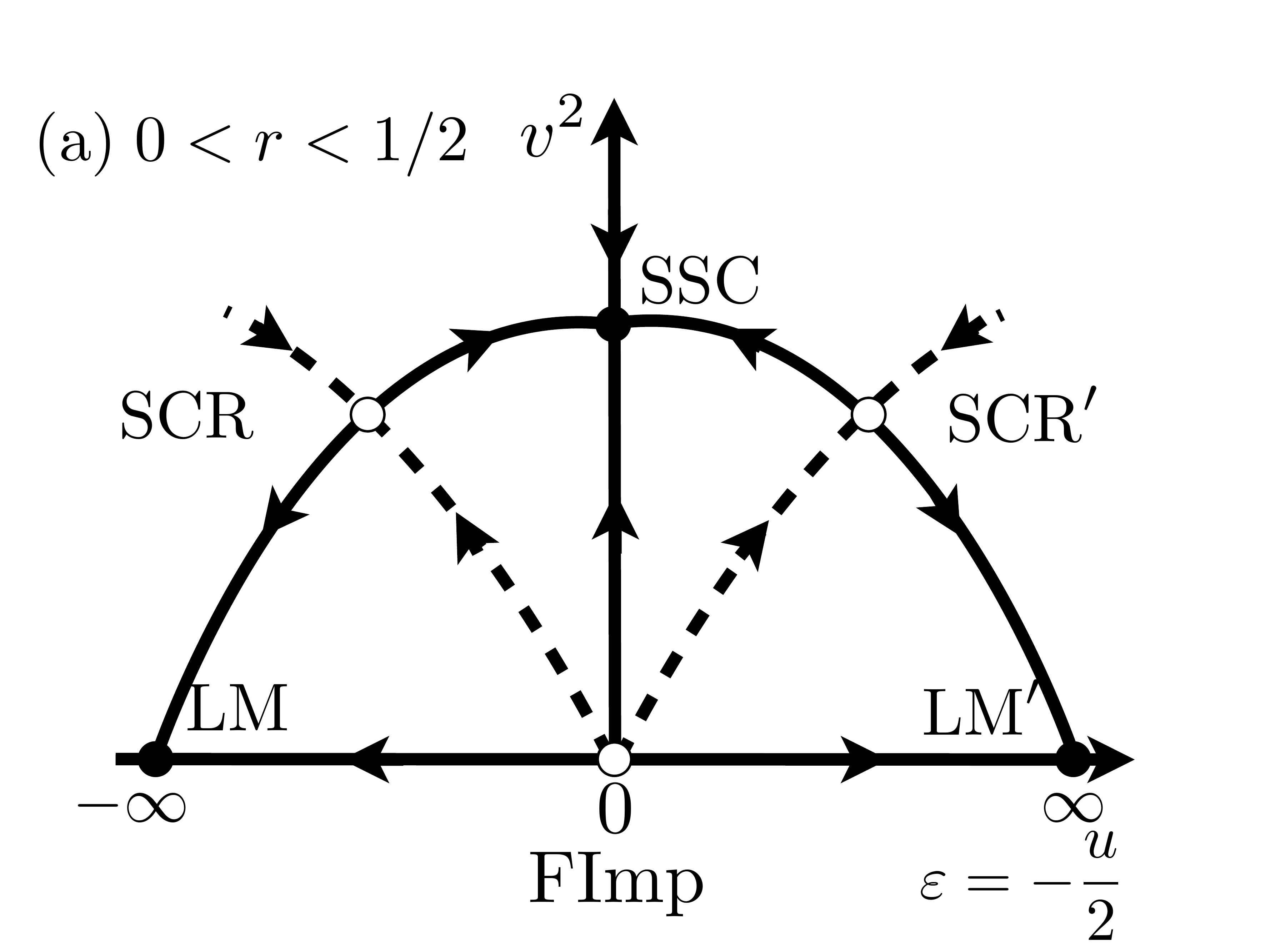}
\includegraphics[width=0.33\textwidth]{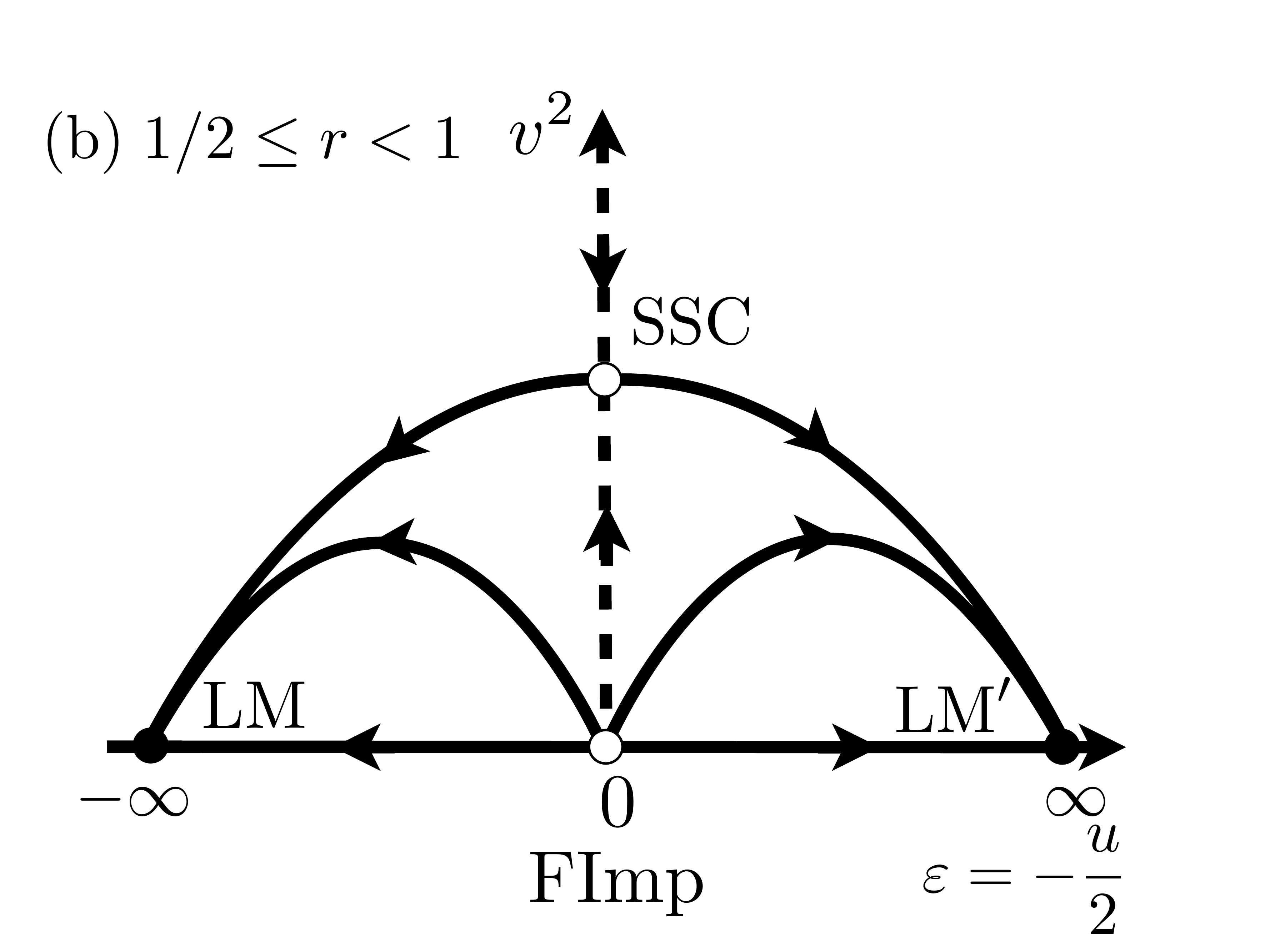}
\includegraphics[width=0.33\textwidth]{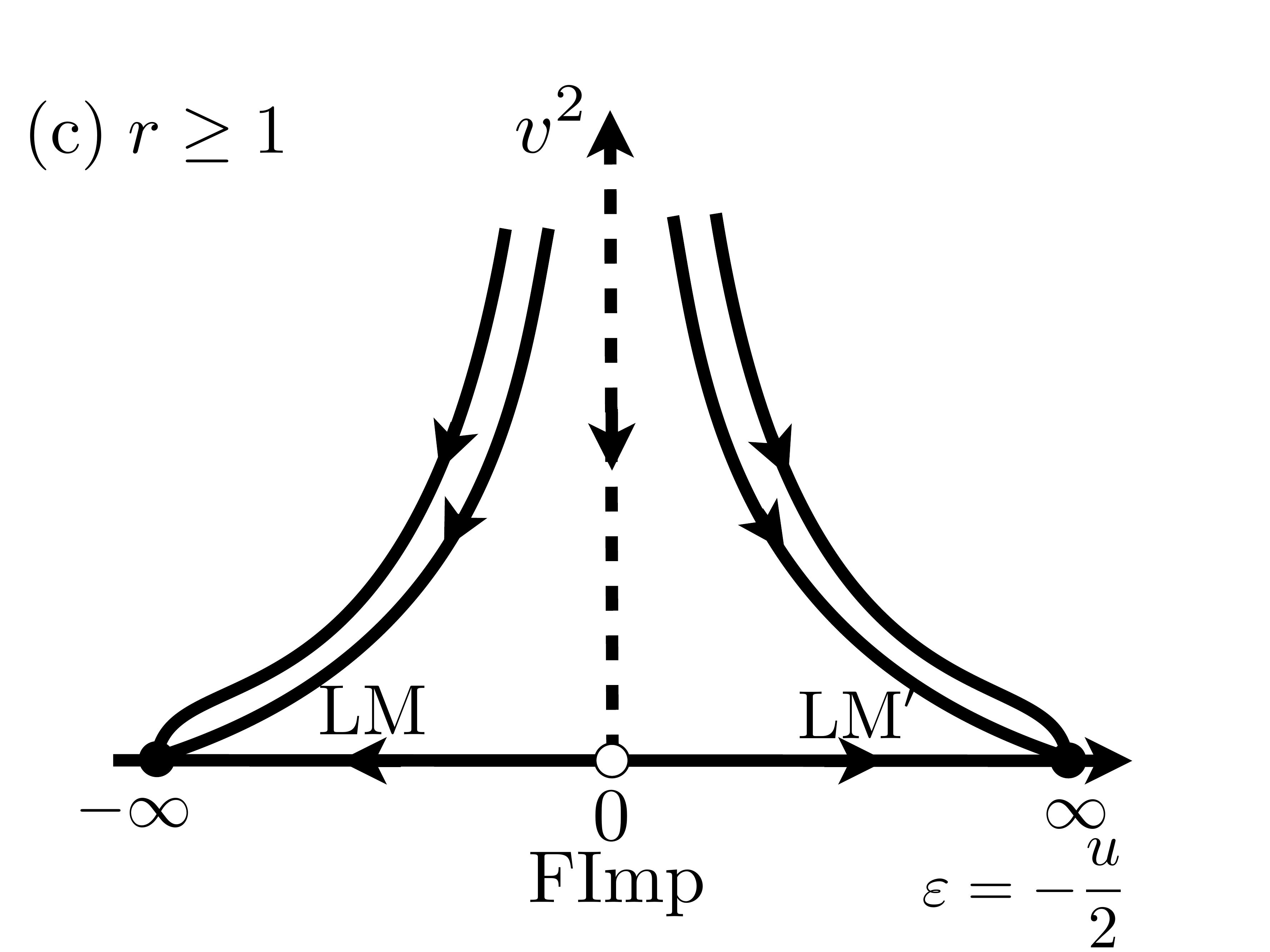}
\caption{
RG flow diagrams for the particle--hole symmetric Anderson model \cite{FV04}, in the
plane spanned by the level energy $\epsilon=-u/2$ and the hybridization $v^2$.
Symbols are as in Fig.~\ref{gonzalezflow}, with LM and LM' corresponding to local moments
formed in the spin or charge channel; the flow of $u$ near SSC is in Eq.~\eqref{eq:rgsym}.
(a) $0<r<1/2$: The critical fixed point SCR (SCR') divides the flow to LM (LM') from that to SSC.
(b) $1/2 \leq r <1$: SCR and SCR' merge with SSC as $r\to1/2^-$, such that SSC is now
unstable.
(c) $r\geq1$: SSC merges with FImp at $\epsilon=v=0$ as $r\to 1^-$.
For all $r\geq1/2$, LM and LM' are the only stable phases in the presence of
particle--hole symmetry. For details see text and Ref.~\cite{FV04}.
}
\label{fig:flowsymmetric}
\end{figure}

A perturbative expansion is now possible in $U$ around the SSC fixed point. The scaling
dimension of the renormalized Coulomb interaction $u$ at SSC is found to be
$\dim[u] = -\overline{r}=-(1-2r)$.
The RG flow of $u$ to two-loop order reads \cite{FV04}
\begin{eqnarray}\label{eq:rgsym}
\frac{du}{d\ln D}=\left(1-2r\right)u-\frac{3\left(\pi-2 \ln 4\right)}{\pi^2}u^3 \;.
\end{eqnarray}
This flow, together with the trivial flow near LM, LM', and FImp is illustrated in
Fig.~\ref{fig:flowsymmetric} \cite{FV04}.
For all $r>0$, LM is a stable fixed point, while SSC is stable only for $r<\rmax$, as can
be seen from Eq.~\eqref{eq:rgsym}. Therefore, a critical fixed point (SCR) emerges for
$0<r<\rmax$, Fig.~\ref{fig:flowsymmetric}b, consistent with Fig.~\ref{gonzalezflow}. Its
properties can now be accessed in a double expansion in $\overline{r}$ and $u$, and
Eq.~\eqref{eq:rgsym} yields for the fixed-point coupling at
${u^*}^2=\frac{\pi^2}{3\left(\pi-2 \ln 4\right)} \overline{r}$. A perturbative
calculation of static critical properties again yields excellent agreement with NRG
results, here for $r\lesssim \rmax$ \cite{FV04}.

Owing to particle--hole symmetry, the behaviour at $\epsilon\geq 0$ is formally identical
to that at $\epsilon\leq 0$, Fig.~\ref{fig:flowsymmetric}, with the latter describing
spin-Kondo physics while the former corresponds to charge-Kondo physics.
Finally, we note that particle--hole symmetry
breaking can be studied perturbatively for small $\overline{r}$, and is found to be
relevant at both SSC and SCR fixed points, again consistent with Fig.~\ref{gonzalezflow}.
In fact, estimating its scaling dimension at SCR as function of $r$, one finds that
particle--hole symmetry breaking changes from being relevant to being irrelevant
once $r$ is reduced below $r\approx0.40$ \cite{lfphd06} -- this value can be identified
with $r^\ast$, Fig.~\ref{gonzalezflow}.

\subsubsection{ACR: Asymmetric Anderson model.}

We now turn to the ACR fixed point present for $r>r^\ast$. This is ultimately relevant for
understanding the Kondo effect in graphene, because there $r=1$ at charge neutrality, and
$t'\neq 0$ breaks particle--hole symmetry already on the level of the band structure.

It was realized in Ref.~\cite{VF04} that the critical theory for ACR is that of a level
crossing of a many-body singlet and a many-body doublet, minimally coupled to conduction
electrons. Using the notation of Ref.~\cite{VF04,FV04}, its Hamiltonian can be written as
\begin{eqnarray}
\label{aiminfu}
\mathcal{H} = \sum_{{\vec{k}},\sigma} \epsilon_{{\vec{k}}} c^\dagger_{{\vec{k}}\sigma}
c^{\phantom{\dagger}}_{{\vec{k}}\sigma} + \epsfb |\sigma\rangle\langle \sigma|
   + g_0 \left[|\sigma\rangle\langle s| c_\sigma(0) + {\rm h.c.}\right]
\end{eqnarray}
where $|\sigma\rangle=|\uparrow\rangle,|\downarrow\rangle$ and $|s\rangle$ represent the three
allowed impurity states.
$\epsfb$ is the tuning parameter (``mass'') of the QPT,
i.e. the (bare) energy difference between doublet and singlet states. The QPT occurs at
some $\epsfb=\epsfc$, with screening present for $\epsfb>\epsfc$.
Remarkably, this theory is identical to a maximally particle--hole asymmetric Anderson
impurity model, Eq.~\eqref{eq:AIM}, where the doubly occupied state has been projected
out, $U\to \infty$, and $(\epsfb,g_0)$ in Eq.~\eqref{aiminfu} have been identified with
$(\epsilon_d,v)$ in Eq.~\eqref{eq:AIM}.

In this model, the point $\epsfb = g_0 = 0$ is dubbed valence-fluctuation fixed point
(VFl). As above, $g_0=0$, $\epsfb=-\infty$ corresponds to LM, while $g_0=0$, $\epsfb=\infty$
describes a fully screened and particle--hole asymmetric singlet state, to be
identified with ASC.

\begin{figure}
\includegraphics[width=0.41\textwidth]{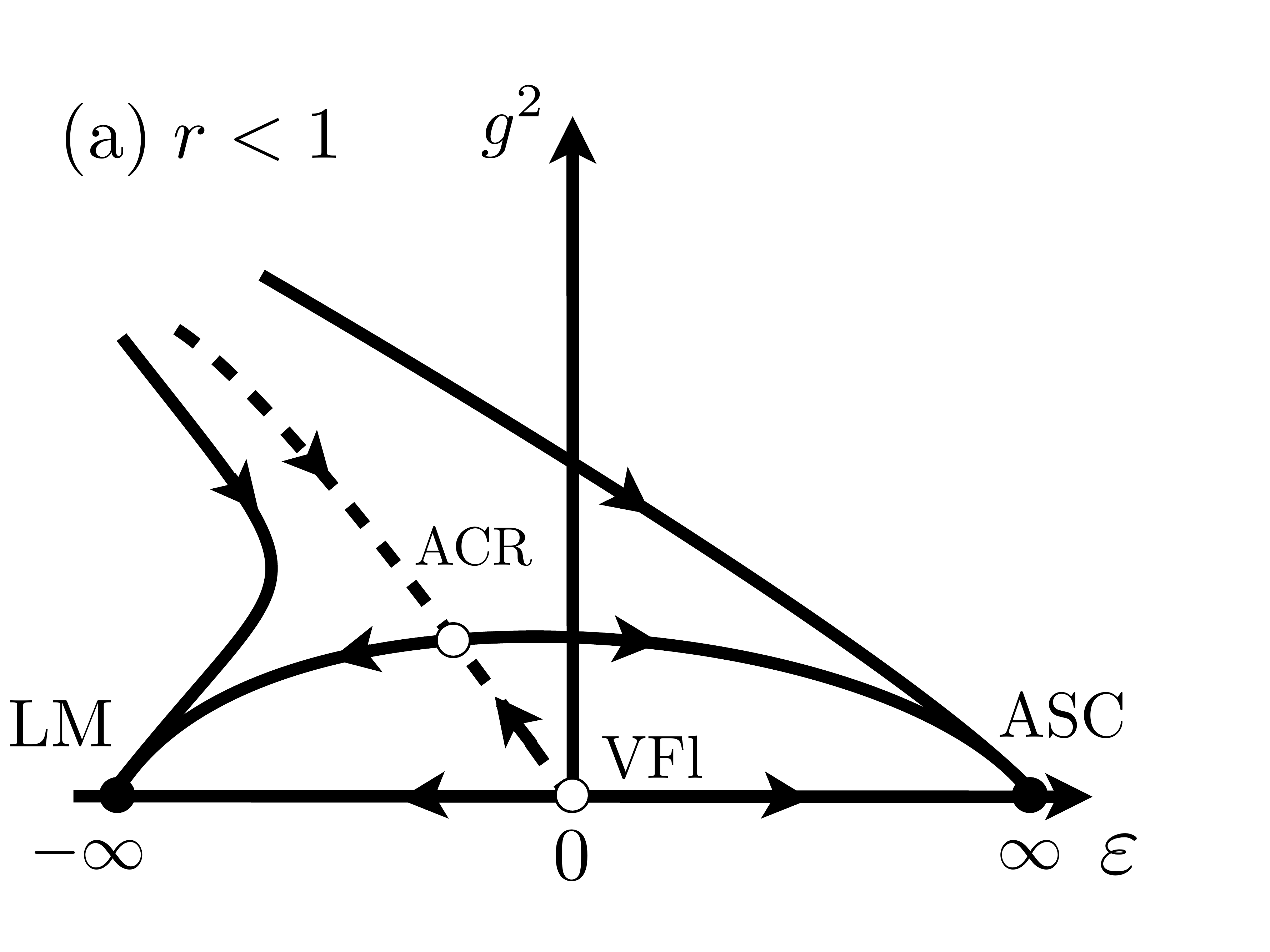}\hspace*{5mm}
\includegraphics[width=0.41\textwidth]{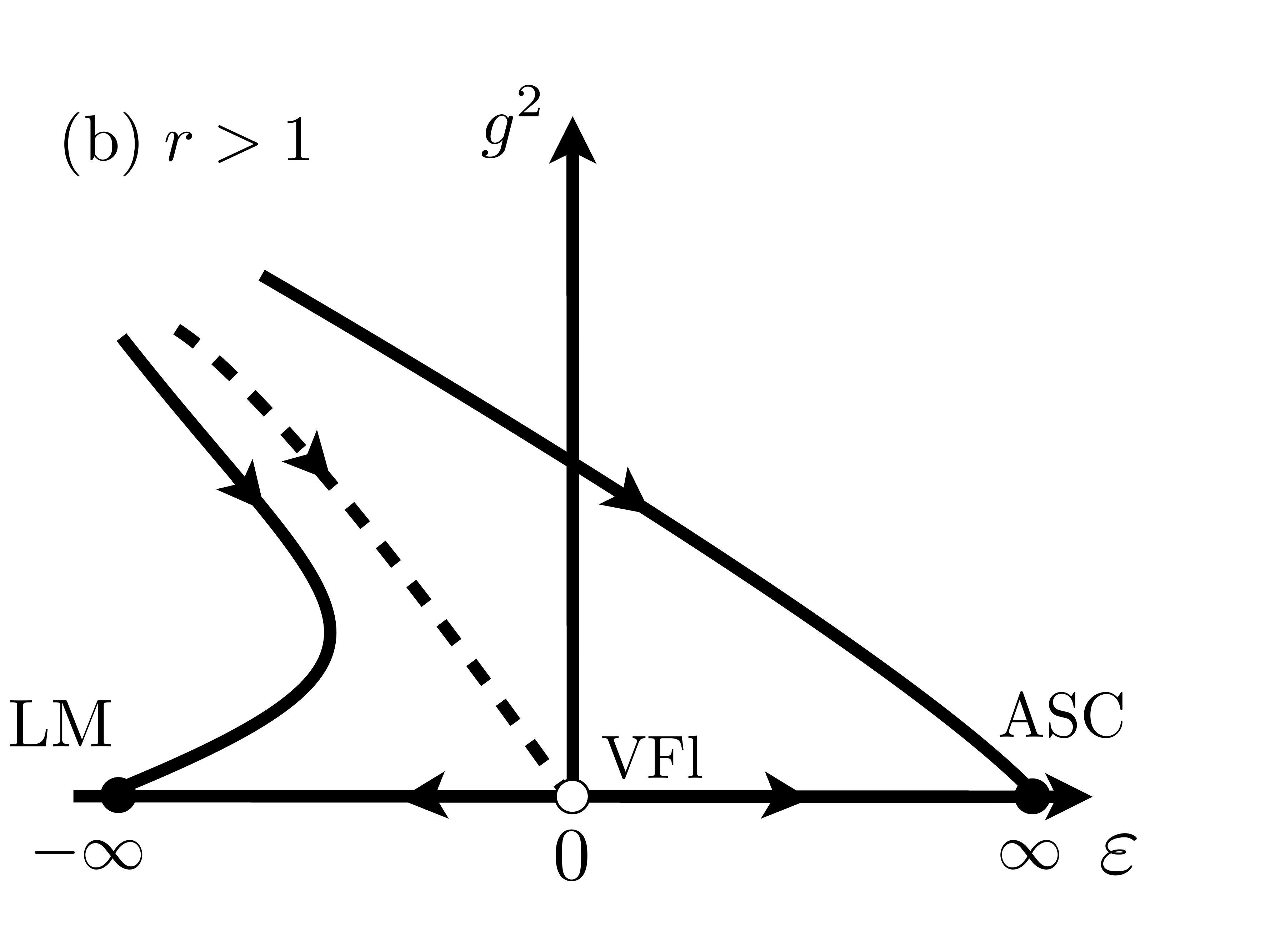}
\caption{
RG flow for the asymmetric Anderson model \cite{VF04,FV04} in the $\epsilon$--$g^2$
plane, obtained from Eq.~\eqref{eq:asyrg}
(a) $r<1$: The critical fixed point ACR separates the flow towards LM from that to ASC.
(b) $r\geq1$: ASC merges with VFl as $r\to 1^-$, which describes a level crossing with
perturbative corrections.
The behaviour near $r=1$ is similar to that of the Landau-Ginzburg model near $d=4$, with
VFl and ASC corresponding to the Gaussian and Wilson-Fisher fixed points, respectively.
}
\label{flowinfu}
\end{figure}

A perturbative expansion is now possible in $g_0$ around VFl. Power counting yields the
scaling dimension of the renormalized hybridization $\dim[g] =
\tilde{r}=\frac{1-r}{2}$. The one-loop flow
equations for $g$ and the renormalized mass $\epsf$ read
\begin{eqnarray}\label{eq:asyrg}
\frac{d g}{d \ln D}&=&-\tilde{r}g+\frac{3}{2}g^3 \nonumber \\
\frac{d \epsf}{d \ln D}&=& -\epsf-g^2+3g^2\epsf \;,
\end{eqnarray}
results to two-loop order can be found in Ref.~\cite{FV04}.
The RG flow is shown in Fig.~\ref{flowinfu} -- this flow has strong similarity to that of
the standard Landau-Ginzburg model. The fact that $g$ is relevant for $r<1$ and irrelevant for
$r>1$ allows us to identify $r=1$ as an upper-critical dimension of the pseudogap
Kondo problem, akin to $d=4$ in the Landau-Ginzburg theory.
For $r<1$ a non-trivial fixed point (ACR) emerges at ${g^*}^2=\frac{2}{3}\tilde{r}$
and $\epsf^*=-\frac{2}{3}\tilde{r}$, Fig.~\ref{flowinfu}a, similar to the celebrated
Wilson-Fisher fixed point. Critical properties, evaluated in a double expansion in
$\tilde{r}$ and $g$, again agree well with NRG results \cite{FV04}.
In contrast, for $r\geq1$ in Fig.~\ref{flowinfu}b, we have ``Gaussian'' behaviour controlled
by the VFl fixed point, which here corresponds to a simple level crossing with
corrections captured by plain perturbation theory in $g_0$.
In the case $r=1$, relevant to charge-neutral graphene, this perturbation theory is
logarithmically divergent at criticality and needs to be resummed, as is standard at the
upper critical dimension.

The structure of the critical theory \eqref{aiminfu} implies that spin and charge
fluctuations are strongly coupled, i.e. suitably defined observables in the charge
sector become critical at the Kondo quantum phase transition controlled by ACR
\cite{VF04,pixley12}.

\subsection{Finite carrier concentration}
\label{sec:dop}

For graphene away from charge neutrality, $\mu\neq 0$, the DOS at the Fermi level is
finite, and consequently a magnetic impurity described by the Kondo model
\eqref{eq:Kondomodel} will be screened in the low-temperature limit for any value of the
Kondo coupling $J_0$ \cite{baskaran07,epl,cornaglia09}.
To be specific, let us consider a Kondo model \eqref{eq:Kondomodel} with bath DOS
\begin{equation}
\rho(\omega) = \frac{1+r}{2D^{r+1}}\,|\omega-\mu|^r\,\Theta(|\w-\mu|-D)\,.
\label{pgdos2}
\end{equation}
In the limit of small $J_0$ the corresponding Kondo temperature will be exponentially small according
to  $\ln\TK \propto -\frac{1}{|\mu|^r}$, but in general $\TK$ needs to be calculated numerically, as
the simple equation \eqref{tkeq} is no longer applicable due to the strong energy
dependence of the DOS.

\begin{figure}[t]
\centering
(a)\includegraphics[width=0.44\textwidth]{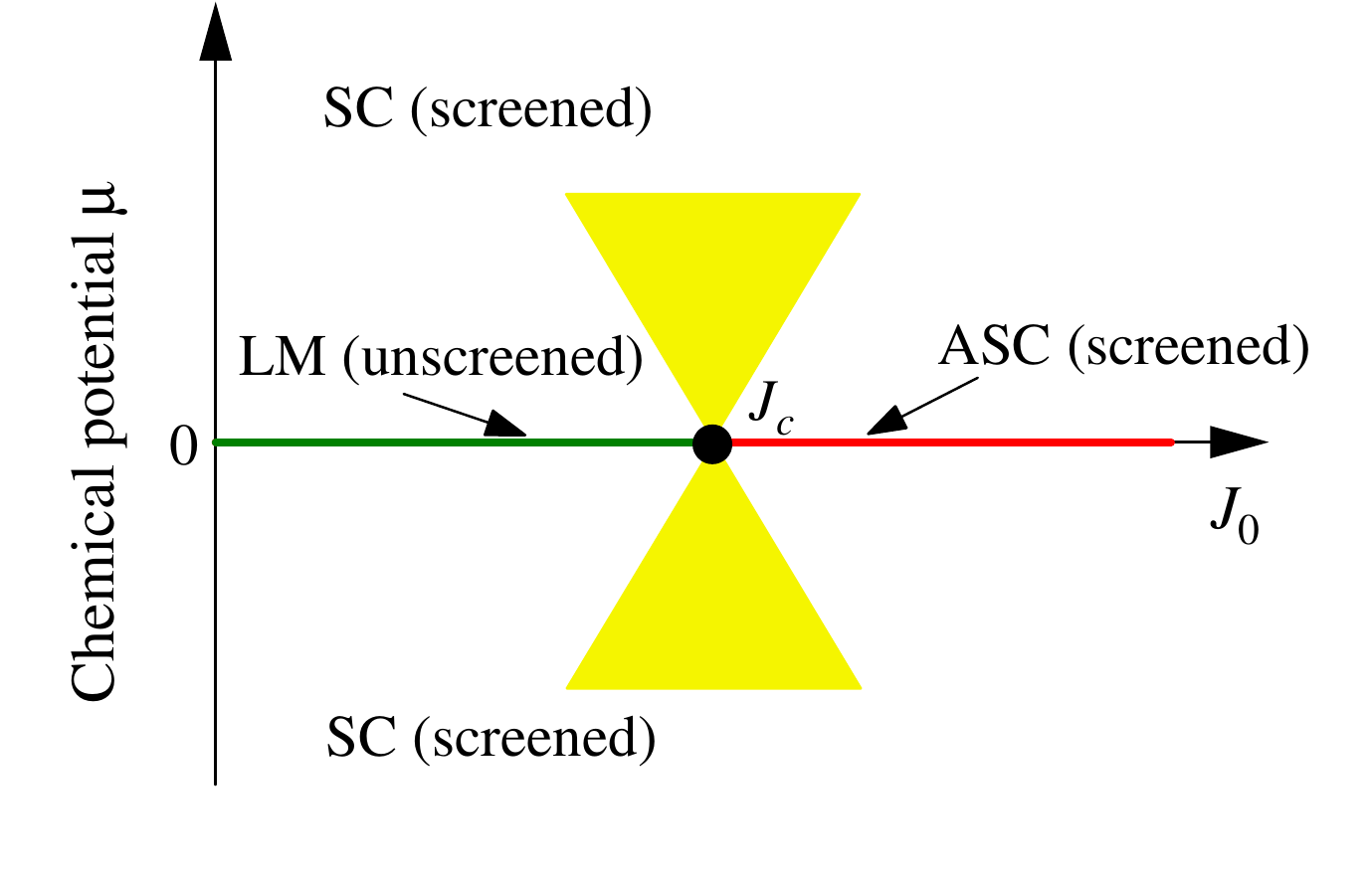}(b)\includegraphics[width=0.52\textwidth]{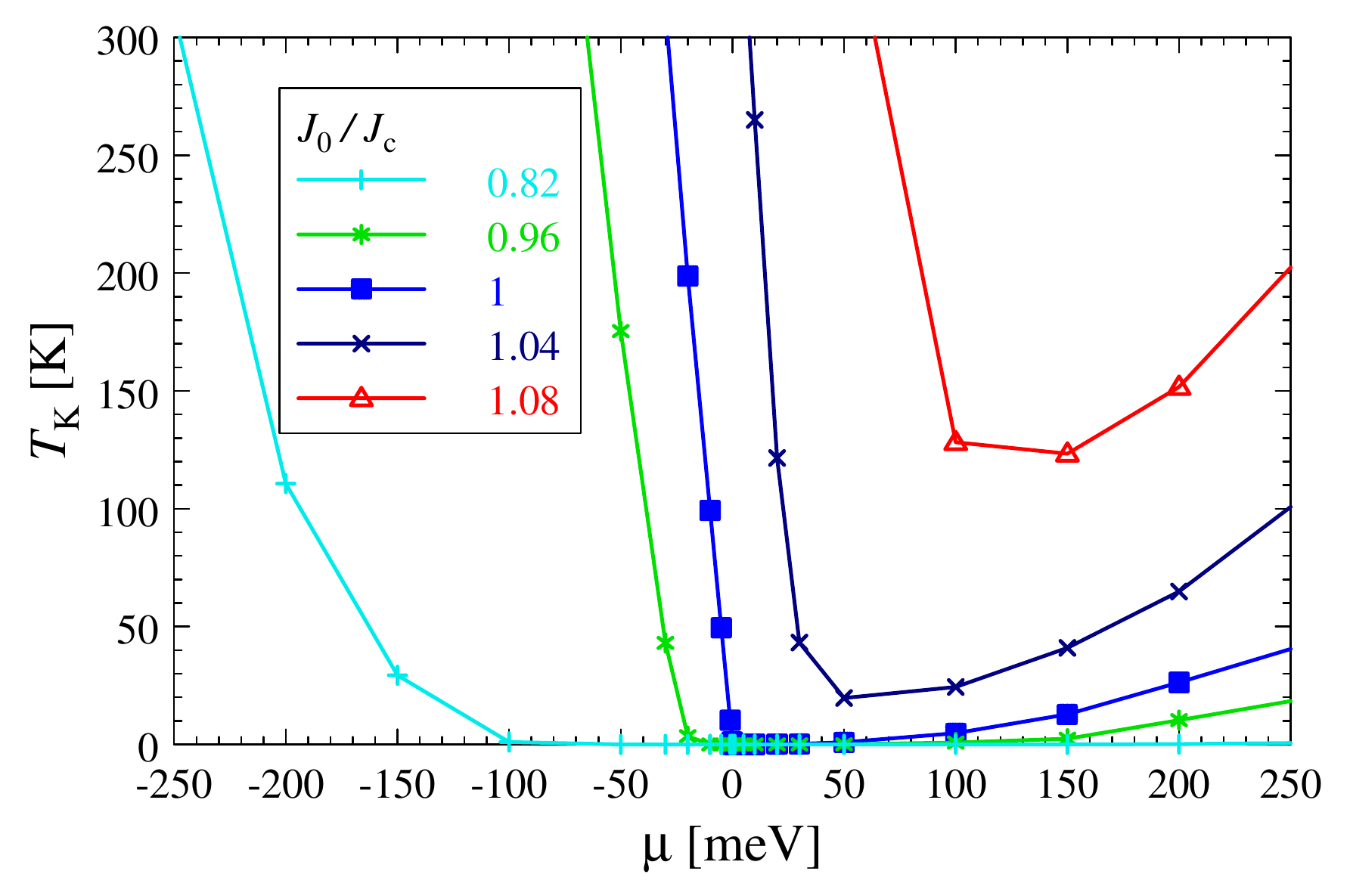}
\caption{
(a) Phase diagram for the pseudogap Kondo problem for $r=1$ in the presence of finite
chemical potential $\mu$. A quantum phase transition between LM and ASC exist in the
charge-neutral case, $\mu=0$, whereas the local moment undergoes Kondo screening for all
$\mu\neq 0$. The shaded region is influenced by quantum criticality, with $\TK(\mu)$
following a power law, for details see text and Ref.~\cite{epl}.
(b) NRG results \cite{epl} for the Kondo temperature $\TK$ as function of $\mu$ for different
values of the Kondo coupling $J_0$, calculated for a DOS appropriate for Co on graphene
\cite{wehling10} where $\Jc \approx 4.3$\,eV.
}
\label{fig:graphene}
\end{figure}

On general grounds, one expects that the presence of the quantum phase transition at
$J_0=\Jc$, $\mu=0$ influences the behaviour at finite $\mu$ as well,
Fig.~\ref{fig:graphene}a. In this quantum critical regime, heuristic scaling arguments
suggest $\TK = \kappa |\mu|$ with a {\em universal} constant $\kappa$ depending on $r$ only.
This problem can be tackled by generalizing the RG equations
\eqref{eq:asyrg} obtained for ACR to a finite chemical potential \cite{epl}:
\begin{eqnarray}\label{rg}
\frac{d g}{d \ln D}&=& -\frac{1-r}{2}g+\frac{g^3}{2}F_1 \left(\frac{\mu}{D} \right)
\nonumber \\
\frac{d \epsf}{d \ln D}&=&-\epsf +g^2 \epsf F_1 \left( \frac{\mu}{D}\right) + g^2 F_2
\left( \frac{\mu}{D}\right)
\end{eqnarray}
with $F_{1,2}(y)=|1+y|^r\pm2|1-y|^r$. The last term in $d \epsf/d\ln D$
describes the level shift due to the real part of the bath Green's function.
A detailed discussion of these equations has been given in Ref.~\cite{epl}, showing that
the asymmetric nature of the critical theory induces a strong asymmetry between the two
signs of $\mu$ in $\TK(\mu)$. In fact, negative $\mu$ drives the near-critical system
directly into the screened phase, whereas positive $\mu$ first induces a crossover to a
spin-1/2 moment which is subsequently screened via a conventional Kondo effect.
For $r<1$, where the ACR fixed point is interacting, the scaling prediction holds with
$\TK = \kappa_\pm |\mu|$ for $\mu\gtrless0$. In contrast, at the upper critical
dimension,
$r=1$, $\TK = \kappa_- |\mu|$ continues to hold for $\mu<0$, while for $\mu>0$
logarithmic corrections and Kondo logarithms conspire such that $\TK \propto |\mu|^x$
where $x\approx 2.6$ is a {\it universal} exponent.

This quantum critical particle--hole asymmetry of $\TK(\mu)$ pertains to the off-critical
situation as well. This is nicely seen in the numerical results,
Fig.~\ref{fig:graphene}b, obtained for a realistic graphene DOS: There is not only
asymmetric behaviour for $J_0\lesssim\Jc$, but also the minimum of $\TK(\mu)$ for $J_0>\Jc$
is {\em not} found at $\mu=0$ but somewhat away from it.

\subsection{Practical consequences for Kondo screening in graphene}
\label{sec:prac}

Based on the analysis presented so far, we can now specify theoretical
predictions for the screening of Kondo impurities in graphene. The half-metallic energy
dependence of the graphene DOS implies that Kondo screening tends to be weaker than in
conventional metals, implying smaller $\TK$, and that the single-parameter scaling known
from the metallic Kondo problem \cite{hewson} will in most cases {\em not} apply.
Such unconventional multi-scale crossovers have been studied explicitly
\cite{epl,cornaglia09}, also in the presence of a quantizing orbital field \cite{dora07}.

The absence of single-parameter scaling may in fact complicate the unambiguous
identification of the Kondo effect -- note that typically the scaling of the
susceptibility or the resistivity correction has been used to argue in favor of Kondo
screening.
We believe that, instead, the doping dependence of $\TK$, the latter
extracted, e.g.,
from the peak width of a tunneling spectrum, can be used as a key indicator of Kondo
screening: $\TK$ will strongly vary with doping, with a minimum near charge neutrality.

A concrete prediction of $\TK$ requires a microscopic modelling for the specific type of
impurity, which would yield the hybridization and interaction terms of an Anderson
impurity model. As discussed in Sec.~\ref{sec:imps}, at present there are considerable
uncertainties in these parameters for all relevant impurities.
Nevertheless, one may use the ab-initio results for the hybridization function of a Co
adatom with $S=1/2$ in an h position \cite{wehling10} (which should apply to NiH as well)
to numerically calculate $\TK$ within an effective Kondo model. Results obtained using
NRG have been presented in Ref.~\cite{epl} and are reproduced in
Fig.~\ref{fig:graphene}b. The strong $\mu$ dependence and the pronounced electron--hole
asymmetry of $\TK(\mu)$ are apparent. At $J_0=\Jc$ the linear and power-law behaviours of
$\TK(\mu)$, advertised in Sec.~\ref{sec:dop}, are nicely visible.
Comments on numbers are in order: (i) Due to the uncertainty in the Coulomb interaction
$U$, the effective Kondo coupling $J_0$ is not known to a good accuracy. Therefore, any
{\em prediction} of $\TK$, in particular near charge neutrality, comes with excessively
large error bars. (ii) In the model leading to Fig.~\ref{fig:graphene}b, we have assumed
SU(2) symmetry. However, according to Ref.~\cite{wehling10}, the Co impurity has an
approximate SU(4) symmetry which is broken down to SU(2) on a scale of 60\,meV.
Therefore, the high-energy flow of the Kondo coupling will differ for the two models,
such that the critical coupling $\Jc$ for Co in this $S=1/2$ state is predicted to be
2.2\,eV, approximately matching the estimate in Ref.~\cite{wehling10}.

Thermodynamic observables like the impurity contributions to susceptibility, entropy, and
specific heat have been calculated for some parameter sets in
Refs.~\cite{wehling10,epl,cornaglia09}, but are difficult to measure in the limit of
dilute impurities.

\begin{figure}[t]
\centering
\includegraphics[width=0.9\textwidth]{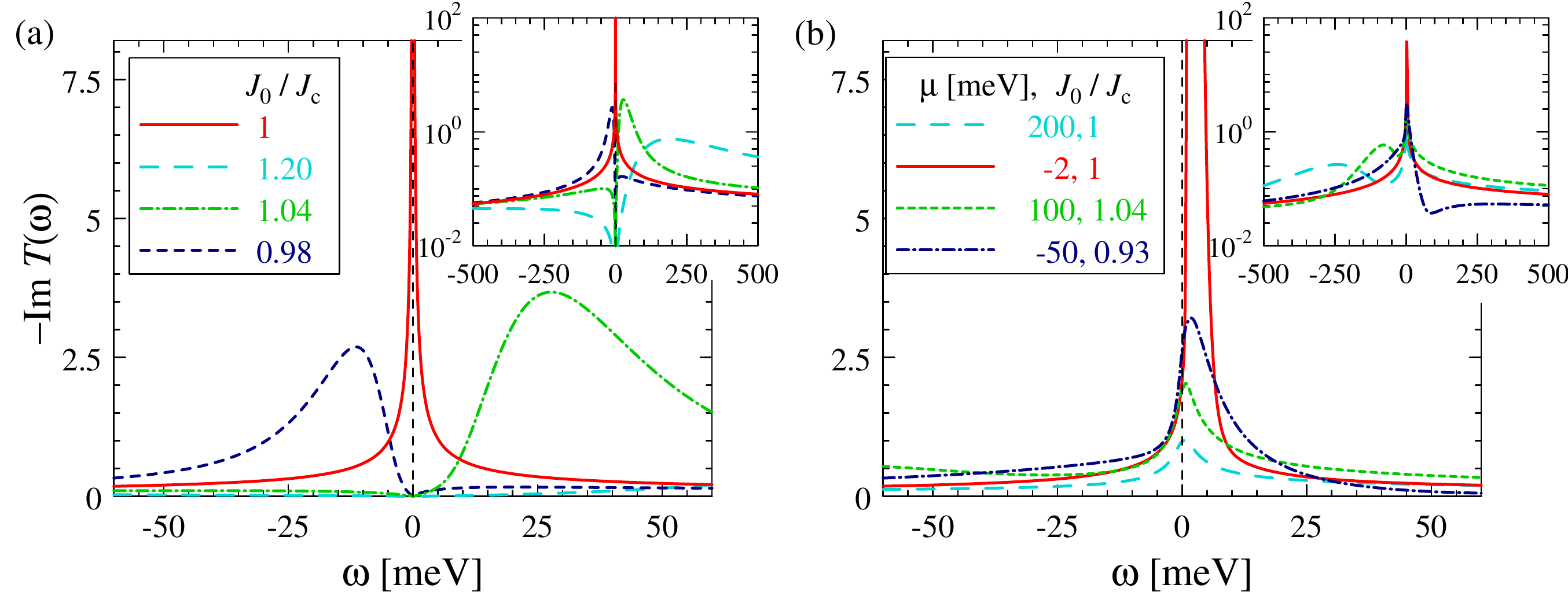}
\caption{
NRG results for the impurity spectral function for different values of the Kondo coupling
$J_0$, calculated for a DOS appropriate for Co on graphene \cite{wehling10} where $\Jc
\approx 4.3$\,eV. (a) Charge-neutral case $\mu=0$. (b) Finite $\mu$; here all parameter
sets yield a $\TK$ between 20 and 30\,K \cite{epl}. The insets show the same data for a
larger energy range and on a logarithmic intensity scale.
}
\label{fig:spec}
\end{figure}

\subsection{STM and quasiparticle interference}

Conduction-electron scattering off impurities can be probed using STM:
Conductance spectra can be recorded at/near the impurity site and as function of the
distance to the impurity. In
particular, spatial variations in the local DOS, $\rho(\vec{r},\w)$, can be interpreted
in terms of impurity-induced energy-dependent Friedel oscillations, so-called
quasiparticle interference (QPI). Analyzing QPI spectra using models of elastic
scattering allows to extract information on both the host band structure and the nature
of the impurity.

For graphene, the initial experiments of Manoharan \cite{manoharan} have triggered a
number of theoretical studies of local spectra
\cite{zhuang09,cornaglia09,wehling10a,saha10,berakdar11} and the expected QPI signal
\cite{wehling10,uchoa09,uchoa11}. A striking feature of the Kondo effect in
charge-neutral graphene is that impurity spectral density is not peaked {\em at} the
Fermi level, but away from it, with vanishing spectral weight at $E_F$
\cite{cornaglia09,VB01}, except for $J_0=J_c$, see Fig.~\ref{fig:spec}a.
In the doped case, single-parameter scaling is again violated
for a large range of parameters \cite{epl,cornaglia09}, but the spectrum returns to being
dominated by a peak near the Fermi level,
Fig.~\ref{fig:spec}b.
We note, however, that a detailed comparison between theory and experiment is lacking to date.


\section{Experiments and open issues}
\label{sec:exp}

Despite numerous attempts to create and study Kondo impurities during several years of
graphene research, the amount of experimental data is still somewhat limited. Both
magnetic adatoms and vacancies have been considered, but clear-cut observations which
unambiguously verify available theories have not been reported to our knowledge.
We start by summarizing the most prominent experiments, and then discuss issues which
might contribute to complicate the interpretation of the data.


\subsection{Experiments}

\subsubsection{Adatoms.}

Isolated Co adatoms placed on top of a graphene sheet have been studied using STM in
Refs.~\cite{manoharan,crommie11}.
Ref.~\cite{manoharan} employed a conducting SiC substrate which shifts the chemical
potential to roughly $\mu=0.25$\,eV. Spectral signatures of Kondo screening were
observed, with $\TK\approx 15$\,K, including the expected splitting of the Kondo peak
upon application of a magnetic field. Surprisingly, $\TK$ was almost identical for Co
atoms placed on the t and h positions of a carbon hexagon, Fig.~\ref{fig:hyb}a. For Co
in the h position, the energy dependence of the small-bias conductance was interpreted
in terms of a two-channel Kondo (2CK) effect, whereas single-channel Kondo (1CK) behaviour
was found for Co in a t position.

In contrast, Ref.~\cite{crommie11} worked with an insulating SiO$_2$ substrate which
allowed for gate tuning of the chemical potential. Here, clear-cut Kondo signatures were
not observed, instead the spectral features mainly reflected charging effects and
vibrational excitations.

\subsubsection{Defects.}

In a second group of experiments, point defects within the graphene sheet were created
either by irradiation \cite{fuhrer11,nair12}, with estimated
defect densities ranging from $10^{-5}$ \cite{fuhrer11} to 0.1 \cite{nair12} per C atom,
or by depositing fluorine adatoms \cite{nair12,hong12}. In the case of irradation, it is
believed that the main defects are carbon vacancies.

\begin{figure}[t]
\center\includegraphics[width=0.6\textwidth]{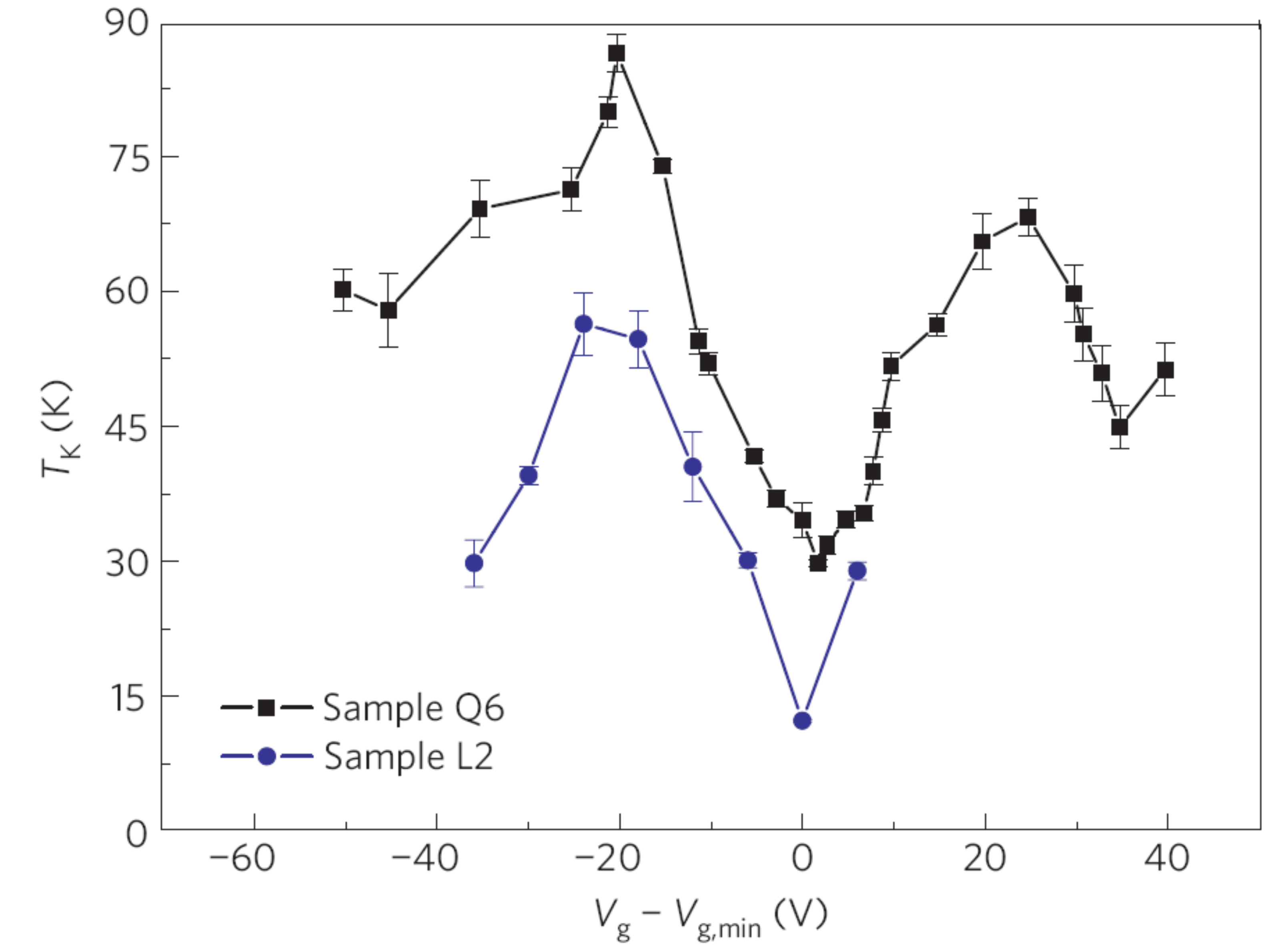}
\caption{
Kondo temperature $\TK$ for vacancy moments in irradiated graphene for two different
samples as function of the gate voltage which controls the carrier concentration
\cite{fuhrer11}; $V_{\rm g,min}$ corresponds to charge-neutral case, i.e. the chemical
potential located at the Dirac point. The $\TK$ values have been extracted from the
temperature dependence of the electrical resistivity (taken from Ref.~\cite{fuhrer11}).
}
\label{fig:fuhrer}
\end{figure}

Ref.~\cite{fuhrer11} studied magnetotransport through an irradiated graphene sample
placed on a Si substrate and found a resistivity increase at low temperatures combined
with negative magnetoresistance, consistent with Kondo screening. Using extensive fits of
$\rho(T)$ to the standard theory of the metallic Kondo effect, $\TK$ was found to vary
between 30 and 90\,K for estimated chemical potentials within [-0.3\,eV,0.3\,eV], see
Fig.~\ref{fig:fuhrer}. Such a $\TK$ variation appears rather small compared to that
expected within a pseudogap Kondo model, see Sec.~\ref{sec:prac} and
Fig.~\ref{fig:graphene}.
Also, it is surprising that the transport data could be fitted to theoretical results for
metallic Kondo screening even near neutrality where standard
single-parameter scaling is not expected due to strong energy dependence of the host DOS.
Despite interesting proposals and ideas \cite{aji11,ogata12}, we feel that a convincing theoretical
explanation for the results of Ref.~\cite{fuhrer11} is missing to date. \footnote{The
Kondo interpretation of the transport data of Ref.~\cite{fuhrer11} has been questioned in
Ref.~\cite{fuhrer_cmt}, where it was instead proposed that electron--electron
interactions in the presence of disorder are responsible for the logarithmic resistivity
increase at low temperature.}

Ref.~\cite{nair12} aimed at quantifying defect-induced magnetism in graphene laminates,
utilizing magnetization measurements away from the dilute limit. For both fluorine
adatoms and vacancies, paramagnetic behaviour of spin-1/2 moments was detected.
The measured magnetic moment per defect was between 0.1 and 0.4\,$\mu_B$ in the vacancy
case, possibly consistent with one spin-1/2 per vacancy in the dilute limit
(due to uncertainties in estimating the vacancy density). For fluorinated graphene the
magnetic moment per adatom was only $10^{-3}\,\mu_B$ -- this extremely small value was
ascribed to adatom clustering, such that only larger clusters contribute one
spin-$\frac{1}{2}$. In these experiments, neither magnetic order nor signatures of Kondo
screening were detected down to 2\,K.

Finally, Ref.~\cite{hong12} investigated weak-localization physics in gated graphene with
fluorine adatoms with dilute concentrations of order $10^{-4}$. The results of the
transport measurements were interpreted in terms of suppressed weak localization due to
spin-flip scattering from fluorine-generated moments, but the Kondo temperature was
estimated to be as small as 0.01\,K for carrier densities of $0.6\times 10^{12}/$cm$^2$.
Such a low $\TK$ would imply a very small magnetic coupling, $J\sim 5$\,meV, between the
fluorine-induced moments and the Dirac electrons of graphene, possibly consistent with
the absence of both magnetic order and screening as measured in Ref.~\cite{nair12}.


\subsection{One-channel vs. two-channel Kondo screening}
\label{sec:pg2ck}

The two-channel Kondo (2CK) effect emerges if a magnetic impurity is coupled
symmetrically to two equivalent screening channels of conduction electrons, such that a
standard Kondo singlet is unstable \cite{nozbl}. Instead, the low-$T$ behaviour is
then governed by a non-trivial intermediate-coupling fixed point with non-Fermi-liquid
properties.

2CK physics being relevant for graphene has been proposed theoretically in
Refs.~\cite{berakdar10,dellanna10,baskaran07}, based on the idea that the electrons in the two
valleys, i.e., near $K$ and $K'$, could form independent screening channels. The
resulting pseudogap 2CK model displays an interesting interplay of pseudogap Kondo
physics and the non-Fermi liquid behaviour of the 2CK effect and has been studied in
Refs.~\cite{GBI,schneider11}.
However, an analysis of possible microscopic models for graphene impurities suggests that
unavoidable inter-valley scattering will invariably couple the two screening channels
such that single-channel Kondo screening prevails at least at low energies and
temperatures \cite{berakdar10}, i.e. below a crossover scale $T_{\rm 1CK}$.
For well localized magnetic moments, inter-valley scattering is strong resulting in
$T_{\rm 1CK}\sim\TK$, such that there is unlikely to be an intermediate regime of 2CK
screening.\footnote{A
conventional single-orbital Anderson model cannot lead to 2CK behavior on general
grounds \cite{tsvelik98}.}
The interpretation of STM data in terms of a 2CK effect in Ref.~\cite{manoharan} is
therefore puzzling.


\subsection{Influence of bulk electron-electron interaction}

The standard analysis of Kondo models assumes non-interacting host electrons, justified
by the assumption of Fermi-liquid behavior and the associated screening of Coulomb
interactions \cite{hewson}. However, in charge-neutral graphene screening is less efficient:
The leading interaction effect is a logarithmic upward renormalization of the Fermi
velocity, such that the DOS is suppressed compared to the non-interacting $|\w|$ result
by a multiplicative logarithm \cite{vozmed94,elias}.
This will lead to a further suppression of the Kondo temperature and to a modification of
the logarithmic corrections at the critical point of the $r=1$ pseudogap Kondo problem
\cite{FV04,epl}, but otherwise not qualitatively alter the behavior.
Away from charge neutrality, the Coulomb interaction is screened, and its effects are minor
similar to standard Fermi liquids.


\subsection{Influence of electron-hole puddles}

A serious complication for quantitative comparisons between theory and experiment, in
particular for graphene sheets near charge neutrality, is the presence of electronic
inhomogeneities, known as electron-hole puddles \cite{martin08,zhang09}. These puddles
can be understood as spatial variations of the local carrier concentration or,
equivalently, of the local Dirac-point energy, with the characteristic length scale (or
puddle size) of 10--20\,nm. While the origin of these puddles is under debate
\cite{zhang09,fehske12,polini12}, a plausible explanation is the influence of charged
defects in the substrate.

In the context of Kondo impurities, these puddles imply a spatial variation of the local
DOS, which itself determines the local Kondo temperature. Thus, a distribution of Kondo
temperatures will be present in a sample with multiple impurities. Importantly, different
macroscopic observables will be dominated by different parts of this distribution:
Whereas the magnetic susceptibility will be dominated by weakly screened (i.e. low-$\TK$)
moments, the electric resistivity receives mainly contributions from strongly screened
(i.e. high-$\TK$) moments. For metallic systems with a broad distribution of Kondo
temperatures, non-Fermi-liquid behaviour may arise, as has been discussed in the framework
of so-called Kondo-disorder models \cite{miranda96,neto98}.

For graphene, an interesting question is whether Kondo disorder could explain the weak
gate-voltage dependence of the transport-$\TK$ in the experiment of Ref.~\cite{fuhrer11}.
However, we believe that this is unlikely to be the case: Weak disorder is insufficient
to significantly modify the $\TK(\mu)$ dependence, whereas strong disorder would yield a
broad $\TK$ distribution which appears incompatible with the fact that the transport data
\cite{fuhrer11} could be well described by the universal metallic Kondo behaviour with a
single $\TK$.


\subsection{Multiple impurities: RKKY interaction and magnetic order}

In samples with a non-vanishing concentration of magnetic impurity moments, the issue of
impurity-induced order becomes relevant, with the coupling between the moments mediated
by conduction electrons via the Ruderman-Kittel-Kasuya-Yosida (RKKY) interaction. For
magnetic moments in metals, magnetic order sets in only at rather high concentrations
(typically 20\% or more), because otherwise Kondo screening is likely to prevail. This
can be different in semiconductors, and indeed semiconductors doped with magnetic ions
display ferromagnetism with Curie temperatures above 100\,K \cite{dms_rev}, possibly
important for spintronics applications.

For graphene, the RKKY interaction has been discussed theoretically
\cite{vozmed05,saremi07,brey07,schaffer10}. For charge-neutral graphene, the oscillations with
distance $r$ typical of RKKY interactions are absent, such that the interaction is
strictly ferromagnetic for moments on the same sublattice and antiferromagnetic for
moments on different sublattices, in both cases falling off as $1/r^3$. Thus, an
inhomogeneous but unfrustrated antiferromagnetic ground state can be expected for
randomly placed moments. Such order is predicted to persist at finite temperature even in
this 2d situation due to the long-range nature of the RKKY interaction
\cite{wessel10}.
Departing from charge neutrality, a crossover to standard metallic behaviour is expected.

The competition of RKKY-induced order with Kondo screening has not been studied in
detail, but it is plausible that most considerations for magnetic moments in metals
apply: Away from charge neutrality, one expects a quantum phase transition between a
Fermi liquid with screened moments and an ordered state which is either antiferromagnetic
or spin-glass-like. As with other quasi-2d metals, the nature of this phase transition is
an open problem \cite{hvl,steglich10,ss12}. In the neutral case, the transition is
between a semimetal and a magnetic insulator -- such a transition has been investigated
in the absence of quenched disorder \cite{herbut06}, but the effect of randomness due to
moment disorder has not been studied.
Experimentally, impurity-induced magnetic order in graphene has not been observed to our
knowledge (see e.g. Ref.~\cite{nair12} for an attempt).


\subsection{Summary}

Studying magnetic moments in graphene holds the prospect of observing exciting phenomena,
such as Kondo physics beyond one-parameter scaling, single-impurity quantum criticality
and associated local non-Fermi-liquid behaviour, magnetic order from dilute moments, and
lattice quantum criticality in strictly two dimensions. While for most of these, a
theoretical framework is available, experimental data is scarce and more experimental
activities are clearly called for.


\section{Outlook: Beyond graphene}
\label{sec:out}

The Kondo effect in graphene is related to a variety of other quantum impurity problems.
This section will highlight the most important connections.

\subsection{Kondo impurities in unconventional superconductors}

The pseudogap Kondo problem was first discussed \cite{withoff} in the context of magnetic
impurities in unconventional superconductors. For BCS states with nodes in the superconducting gap
function, the density of states of Bogoliubov quasiparticles vanishes algebraically as
$\w\to 0$. Moreover, a locally coupled impurity, $V_k=V$ in Eq.~\eqref{eq:AIM}, is not
directly influenced by pairing, because the local anomalous Green's function vanishes in
an unconventional superconductor \cite{FV05}. Then, the Kondo problem in such a
superconductor is equivalent to the pseudogap Kondo problem described in
Sec.~\ref{sec:pg}, with a $|\w|^r$ density of states at low energies and $r=1$ ($r=2$)
for $d$-wave ($p$-wave) superconductors. Importantly, the Fermi energy is always pinned
to charge neutrality, i.e. the crossovers described in Sec.~\ref{sec:dop} cannot be
accessed.

A clear-cut experimental observation of Kondo screening in unconventional superconductors
ideally requires $\TK\lesssim\Tc$, with $\Tc$ being the superconducting transition
temperature, together with the ability to tune either the host gap or the Kondo coupling.
While the latter can be efficiently varied in nanostructures with quantum dots, we are
not aware of corresponding experimental results, likely because of the lack of suitable
unconventional superconductors as lead materials.
In bulk superconductors with impurities, signatures of Kondo screening have been
detected in NMR measurements on Zn-doped high-$\Tc$ cuprates of the \ybco\ family
\cite{bobroff99,bobroff01}. Similarly, STM experiments on \bscco\ have detected large
low-energy conductance peaks near Zn impurities \cite{seamuszn}, which subsequently have
been interpreted in terms of a Kondo resonance \cite{tolya01,VB01}.\footnote{For cuprates, it has
been assumed that the non-magnetic Zn ion induces a magnetic moment by a mechanism of
dimer breaking, as happens in spin-gapped magnets such as spin ladders.}
It should be noted, however, that other interpretations for the STM data have been put
forward as well \cite{seamuszn,interlayer,bala_rmp}, and a concise picture for impurity
effects in cuprates has not yet emerged.


\subsection{Kondo impurities on the surface of topological insulators}

The 2d surface states of 3d topological insulators (TI) admit a low-energy
description in terms of a Dirac equation. The resulting electronic properties are
therefore similar to that of graphene, with a few important differences: (i) there is a single
Dirac cone (or, more generally, an odd number) per surface, and (ii) the role of the
pseudospin (or sublattice) in the graphene case is taken by the physical spin, such that
TI surface states display spin-momentum locking, and there is no additional spin
degeneracy.

The physics of Kondo impurities in this setting has been analyzed theoretically in a
number of papers recently \cite{feng10,zitko10,kim10}. The main conclusion is that,
despite the non-trivial topological structure of the TI surface states, the corresponding
local Kondo problem for a spin $S=1/2$ impurity can be mapped onto the standard
pseudogap Kondo model of Sec.~\ref{sec:pg}. Consequently, strong deviations from
conventional, i.e. metallic, Kondo screening are expected once the Fermi level is tuned
close to the Dirac-point energy of the surface states.
Experimentally, magnetic moments on TI surfaces have been investigated in a few papers
\cite{kimura12,rader12,honolka12}, but not in the dilute regime with focus on Kondo screening.


\subsection{Kondo physics in spin liquids}

A somewhat different, but still overlapping, topic is the physics of magnetic impurities
embedded in quantum magnets. The closest relation to the graphene Kondo problem is found
for host magnets without semiclassical long-range order, but still a gapless spectrum of
excitations. Two interesting cases will be discussed in the following.

In quantum-critical magnets, located near a zero-temperature transition between an
antiferromagnet and a paramagnet, the elementary host excitations are spin-1 critical
magnons which interact with the impurity spin via a Yukawa-type (i.e. three-point)
coupling. As a result, true Kondo singlet formation is not possible. In dimensions
$1<d<3$, RG studies have predicted a partial screening of the spin, described by a
non-trivial intermediate-coupling fixed point \cite{sbv99,sv03}. In $d=2$ this prediction
has been verified numerically \cite{sandvik07}.

Magnetic impurities in gapless spin-liquid phases are more akin to Kondo problems: Here
the host excitations are typically spin-1/2 spinons coupled to a U(1) and Z$_2$ gauge
field. The spinons couple to the impurity spin with a Kondo-like (i.e. four-point)
interaction. The physics depends on the nature of the spinons, and a few cases have been
discussed in the literature.
Linearly dispersing bosonic spinons yield a rich phase diagram, with a variety of
possible $T=0$ phases, including the possibility of full Kondo screening, and quantum
phase transitions \cite{FFV06}.
In contrast, fermionic spinons lead to physics similar to standard Kondo expectations: In
the presence of a spinon Fermi surface, the impurity spin gets always screened at low $T$
\cite{ribeiro12}. In the case of 2d Dirac spinons of an algebraic spin liquid, a quantum
phase transition not unlike that of the pseudogap Kondo problem, Sec.~\ref{sec:pg},
emerges \cite{cassa,kim08,dhochak10}. However, it should be noted that the influence of
gauge fields beyond perturbation theory has been neglected in the published treatments.


\ack

We thank A. Castro Neto, P. Coleman, P. Cornaglia, K. Ingersent, H. Manoharan, A. Mitchell, A. Rosch, B. Uchoa, and T. Wehling
for illuminating discussions as well as
F. B. Anders, R. Bulla, S. Florens, M. Kircan, A. Polkovnikov, S. Sachdev, I. Schneider, and M. Zwiebler
for collaborations on these and related topics.
This research was supported by the DFG via SFB 608, SFB/TR 12, FOR 960, GRK 1621, and the
Emmy-Noether programme (FR 2627/3-1).



\section*{References}
\begin{thebibliography}{100}

\bibitem{kondo}
Kondo J 1964
{\em Prog. Theor. Phys.} {\bf 32} 37

\bibitem{hewson} Hewson A C 1997
{\em The Kondo Problem to Heavy Fermions}
(Cambridge: Cambridge University Press).


\bibitem{novo1}
Novoselov K S, Geim A K, Morozov S V, Jiang D, Zhang Y, Dubonos S V, Grigorieva I V and Firsov A A 2004
{\em Science} {\bf 306} 666

\bibitem{novo2}
Novoselov K S, Geim A K, Morozov S V, Jiang D, Katsnelson M I, Grigorieva I V, Dubonos S V and Firsov A A 2005
{\em Nature} {\bf 438} 197

\bibitem{neto_rmp}
Castro Neto A H, Guinea F, Peres N M R, Novoselov K S and Geim A K 2009
{\em Rev. Mod. Phys.} {\bf 81} 109

\bibitem{sarma_rmp}
Das Sarma S, Adam S, Hwang E H and Rossi E 2011
{\em Rev. Mod. Phys.} {\bf 83} 407

\bibitem{mv_rev}
Vojta M 2006
{\em Phil. Mag.} {\bf 86} 1807

\bibitem{wallace47}
Wallace P R 1947
{\em Phys. Rev.} {\bf{71}} 622


\bibitem{Anderson61}
Anderson P W 1961
{\em Phys. Rev.} {\bf 124} 41

\bibitem{eigler00}
Manoharan H C, Lutz C P and Eigler D M 2000
{Nature} {\bf{403}} 512


\bibitem{Costi09}
Costi T A, Bergqvist L, Weichselbaum A, von Delft J, Micklitz T, Rosch A, Mavropoulos P, Dederichs P H, Mallet F, Saminadayar L, Bauerle C 2009
{\em Phys. Rev. Lett.} {\bf 102} 056802


\bibitem{wehling10}
Wehling T O, Balatsky A V, Katsnelson M I, Lichtenstein A I and Rosch A 2010
{\em Phys. Rev. B} {\bf 81} 115427

\bibitem{uchoa09}
Uchoa B, Yang L, Tsai S-W, Peres N M R and Castro Neto A H 2009
{\em Phys. Rev. Lett.} {\bf 103} 206804

\bibitem{berakdar10}
Zhu Z-G, Ding K-H and Berakdar J 2010
{\em EPL} {\bf 90} 67001

\bibitem{uchoa11}
Uchoa B, Yang L, Tsai S-W, Peres N M R and Castro Neto A H 2011
{\em preprint} arXiv:1105.4893

\bibitem{uchoa11b}
Uchoa B, Rappoport T G and Castro Neto A H 2011
{\em Phys. Rev. Lett.} {\bf 106} 016801;
{\em Phys. Rev. Lett.} {\bf 106} 159901(E)

\bibitem{wehling11}
Wehling T O, Lichtenstein A I and Katsnelson M I 2011
{\em Phys. Rev. B} {\bf 84} 235110

\bibitem{kotliar10}
Jacob D and Kotliar G 2010
{\em Phys. Rev. B} {\bf 82}, 085423

\bibitem{rudenko12}
Rudenko A N, Keil F J, Katsnelson M I and Lichtenstein A I 2012
{\em preprint} arXiv:1206.1222

\bibitem{uchoa08}
Uchoa B, Kotov V N, Peres N M R and Castro Neto A H 2008
{\em Phys. Rev. Lett.} {\bf 101} 026805

\bibitem{GBI}
Gonzalez-Buxton C and Ingersent K 1998
{\em Phys. Rev. B} {\bf 57} 14254


\bibitem{grigorieva12}
Nai R R, Sepioni M, Tsai I-L, Lehtinen O, Keinonen J, Krasheninnikov A V, Thomson T, Geim A K and Grigorieva I V 2012
{\em Nature Phys.} {\bf 8} 199

\bibitem{kawakami12}
McCreary K M, Swartz A G, Han W, Fabian J and Kawakami R K 2012
{\em preprint} arXiv:1206.2628


\bibitem{pereira06}
Pereira V M, Guinea F, Lopes dos Santos J M B, Peres, N M R and Castro Neto A H 2006
{\em Phys. Rev. Lett.} {\bf 96} 036801

\bibitem{helm06}
Yazyev O V and Helm L 2007
{\em Phys. Rev. B} {\bf 75} 125408

\bibitem{pruschke11}
Haase P, Fuchs S, Pruschke T, Ochoa H and Guinea F 2011
{\em Phys. Rev. B} {\bf 83} 241408(R)

\bibitem{balseiro12}
Sofo J O, Usaj G, Cornaglia P S, Suarez A M, Hernandez-Nieves A D and Balseiro C A 2012
{\em Phys. Rev. B} {\bf 85} 115405

\bibitem{yndurain12}
Palacios J J and Yndurain F 2012
{\em Phys. Rev. B} {\bf 85} 245443

\bibitem{cazalilla12}
Cazalilla M A, Iucci A, Guinea F and Castro Neto A H 2012
{\em preprint} arXiv:1207.3135

\bibitem{VB02}
Vojta M and Bulla R 2002
{\em Eur. Phys. J B} {\bf 28} 283


\bibitem{withoff}
Withoff D and Fradkin E 1990
{\em Phys. Rev. Lett.} {\bf 64} 1835

\bibitem{hentschel07}
Hentschel M and Guinea F 2007
{\em Phys. Rev. B} {\bf 76} 115407

\bibitem{bulla97}
Bulla R, Pruschke T and Hewson A C 1997
{\em J. Phys.: Condens. Matter} {\bf 9} 10463

\bibitem{bulla00}
Bulla R, Glossop M T, Logan D E and Pruschke T 2000
{\em J. Phys.: Condens. Matter} {\bf 12} 4899

\bibitem{si02}
Ingersent K and Si Q 2002
{\em Phys. Rev. Lett.} {\bf 89} 076403

\bibitem{VF04}
Vojta M and Fritz L 2004
{\em Phys.~Rev.~B} {\bf 70} 094502

\bibitem{FV04}
Fritz L and Vojta M 2004
{\em Phys.~Rev.~B} {\bf 70} 214427

\bibitem{readnewns}
Read N and Newns D M 1983
{\em J. Phys. C} {\bf 16} 3273

\bibitem{si98}
Ingersent K and Si Q 1998
{\em preprint} arXiv:cond-mat/9810226
(unpublished)

\bibitem{cassa}
Cassanello C R and Fradkin E 1996
{\em Phys. Rev. B} {\bf 53} 15079

\bibitem{tolya01}
Polkovnikov A, Sachdev S and Vojta M 2001
{\em Phys. Rev. Lett.} {\bf 86} 296

\bibitem{tolya02}
Polkovnikov A 2002
{\em Phys. Rev. B} {\bf 65} 064503

\bibitem{zhuang09}
Zhuang H B, Sun Q-f and Xie X C 2009
{\em EPL} {\bf 86} 58004

\bibitem{dellanna10}
Dell'Anna L 2010
{\em J. Stat. Mech.} P01007

\bibitem{FFV06}
Fritz L, Florens S and Vojta M 2006
{\em Phys. Rev. B} {\bf 74} 144410

\bibitem{glossop11}
Glossop M T, Kirchner S, Pixley J H and Si Q 2011
{\em Phys. Rev. Lett.} {\bf 107} 076404

\bibitem{pixley12}
Pixley J H, Kirchner S, Ingersent K and Si Q 2011
{\em preprint} arXiv:1108.5227

\bibitem{lfphd06}
Fritz L 2006
{\em PhD thesis} Universit\"at Karlsruhe

\bibitem{baskaran07}
Sengupta K and Baskaran G 2007
{\em Phys.~Rev.~B} {\bf 77} 045417

\bibitem{epl}
Vojta M, Fritz L and Bulla R 2010
{\em EPL} {\bf 90} 27006

\bibitem{cornaglia09}
Cornaglia P S, Usaj G and Balseiro C A 2009
{\em Phys. Rev. Lett.} {\bf 102} 046801

\bibitem{dora07}
Dora B and Thalmeier P 2007
{\em Phys. Rev. B} {\bf 76} 115435


\bibitem{manoharan}
Manoharan H C 2011
{\em Bulletin of the American Physical Society} BAPS.2011.MAR.P2.1
(unpublished)


\bibitem{wehling10a}
Wehling T O, Dahal H P, Lichtenstein A I, Katsnelson M I, Manoharan H C and Balatsky A V 2010
{\em Phys. Rev. B} {\bf 81} 085413

\bibitem{saha10}
Saha K, Paul I and Sengupta K 2010
{\em Phys. Rev. B} {\bf 81} 165446

\bibitem{berakdar11}
Zhu Z-G and Berakdar J 2011
{\em Phys. Rev. B} {\bf 84} 165105

\bibitem{VB01}
Vojta M and Bulla R 2001
{\em Phys. Rev. B} {\bf 65} 014511


\bibitem{crommie11}
Brar V W, Decker R, Solowan H-M, Wang Y, Maserati L, Chan K T, Lee H, Girit C O, Zettl A,
Louie S G, Cohen M L and Crommie M F 2011
{\em Nature Phys.} {\bf 7} 43

\bibitem{fuhrer11}
Chen J-H, Li L, Cullen W G, Williams E D and Fuhrer M S 2011
{\em Nature Phys.} {\bf 7} 535

\bibitem{nair12}
Nair R R, Sepioni M, Tsai I-L, Lehtinen O, Keinonen J, Krasheninnikov A V, Thomson T, Geim A K and
Grigorieva I V 2012
{\em Nature Phys.} {\bf 8} 199

\bibitem{hong12}
Hong X, Zou K, Wang B, Cheng S-H and Zhu J 2012
{\em Phys. Rev. Lett.} {\bf 108} 226602


\bibitem{aji11}
Chao S-P and Aji V 2011
{\em Phys. Rev. B} {\bf 83} 165449

\bibitem{ogata12}
Kanao T, Matsuura H and Ogata M 2012
{\em J. Phys. Soc. Jpn.} {\bf 81} 063709


\bibitem{fuhrer_cmt}
Jobst H and Weber H B 2012
{\em Nature Phys.} {\bf 8} 352;
Chen J-H {\em et al.} 2012
{\em Nature Phys.} {\bf 8} 353


\bibitem{nozbl}
Nozi\`eres P and Blandin A 1980
{\em J.~de Physique} {\bf 41} 193

\bibitem{schneider11}
Schneider I, Fritz L, Anders F B, Benlagra A and Vojta M 2011
{\em Phys. Rev. B} {\bf 84} 125139

\bibitem{tsvelik98}
Coleman P and Tsvelik A M 1998
{\em Phys. Rev. B} {\bf 59} 12757


\bibitem{vozmed94}
Gonzalez J, Guinea F and Vozmediano M A H 1994
{\em Nucl. Phys. B} {\bf 424} 595

\bibitem{elias}
Elias D C, Gorbachev R V, Mayorov A S, Morozov S V, Zhukov A A,
Blake P, Ponomarenko L A, Grigorieva I V, Novoselov K S, Guinea F and Geim A K 2011
{\em Nature Phys.} {\bf 7} 701


\bibitem{martin08}
Martin J, Akerman N, Ulbricht G, Lohmann T, Smet J H, von Klitzing K and Yacoby A 2008
{\em Nature Phys.} {\bf 4} 144

\bibitem{zhang09}
Zhang Y, Brar V W, Girit C, Zettl A and Crommie M F 2009
{\em Nature Phys.} {\bf 5} 722

\bibitem{fehske12}
Schubert G and Fehske H 2012
{\em Phys. Rev. Lett.} {\bf 108} 066402

\bibitem{polini12}
Gibertini M, Tomadin A, Guinea F, Katsnelson M I and Polini M 2012
{\em Phys. Rev. B} {\bf 85} 201405(R)

\bibitem{miranda96}
Miranda E, Dobrosavljevic V and Kotliar G 1996
{\em J. Phys. Cond. Matter} {\bf 8} 9871

\bibitem{neto98}
Castro Neto A H and Jones B 2000
{\em Phys. Rev. B} {\bf 62} 14975


\bibitem{dms_rev}
Jungwirth T, Sinova J, Malek J, Kucera J and MacDonald A H 2006
{\em Rev. Mod. Phys.} {\bf 78} 809

\bibitem{vozmed05}
Vozmediano M A H, Lopez-Sancho M P, Stauber T and Guinea F 2005
{\em Phys. Rev. B} {\bf 72} 155121

\bibitem{saremi07}
Saremi S 2007
{\em Phys. Rev. B} {\bf 76} 184430

\bibitem{brey07}
Brey L, Fertig H A and Das Sarma S 2007
{\em Phys. Rev. Lett.} {\bf 99} 116802

\bibitem{schaffer10}
Black-Schaffer A M 2010
{\em Phys. Rev. B} {\bf 80} 205416

\bibitem{wessel10}
Fabritius T, Laflorencie N and Wessel S 2010
{\em Phys. Rev. B} {\bf 82} 035402

\bibitem{hvl}
von L\"ohneysen H, Rosch A, Vojta M  and W\"olfle P 2007
{\em Rev. Mod. Phys.} {\bf 79} 1015

\bibitem{steglich10}
Gegenwart P, Si Q and Steglich F 2008
{\em Nature Phys.} {\bf 4}, 186

\bibitem{ss12}
Sachdev S, Metlitski M A and Punk M 2012
{\em Journal of Physics: Cond. Matter} {\bf 24}, 294205

\bibitem{herbut06}
Herbut I F 2006
{\em Phys. Rev. Lett.} {\bf 97} 146401


\bibitem{FV05}
Fritz L and Vojta M 2005
{\em Phys. Rev. B} {\bf 72} 212510

\bibitem{bobroff99}
Bobroff J, MacFarlane W A, Alloul H, Mendels P, Blanchard N, Collin G and Marucco J-F 1999
{\em Phys. Rev. Lett.} {\bf 83} 4381

\bibitem{bobroff01}
Bobroff J, Alloul H, MacFarlane W A, Mendels P, Blanchard N, Collin G and Marucco J-F 2001
{\em Phys. Rev. Lett.} {\bf 86} 4116

\bibitem{seamuszn}
Pan S H, Hudson E W, Lang K M, Eisaki H, Uchida S and Davis J C 2000
{\em Nature} {\bf 403} 746

\bibitem{interlayer}
Martin I, Balatsky A V and Zaanen J 2002
{\em Phys. Rev. Lett.} {\bf 88}, 097003

\bibitem{bala_rmp}
Balatsky A V, Vekhter I and Zhu J-X 2006
{\em Rev. Mod. Phys.} {\bf 78} 373


\bibitem{feng10}
Feng X-Y, Chen W-Q, Gao J-H, Wang Q-H and Zhang F C 2010
{\em Phys. Rev. B} {\bf 81} 235411

\bibitem{zitko10}
Zitko R 2010
{\em Phys. Rev. B} {\bf 81} 241414

\bibitem{kim10}
Tran M-T and Kim K-S 2010
{\em Phys. Rev. B} {\bf 82} 155142

\bibitem{kimura12}
Ye M, Eremeev S V, Kuroda K , Krasovskii E E, Chulkov E V, Takeda Y, Saitoh Y, Okamoto K, Zhu S Y, Miyamoto K,
Arita M, Nakatake M, Okuda T, Ueda Y, Shimada K, Namatame H, Taniguchi M and Kimura A 2012
{\em Phys. Rev. B} {\bf 85} 205317

\bibitem{rader12}
Scholz M R, Sanchez-Barriga J, Marchenko M, Varykhalov A, Volykhov A, Yashina L V and Rader O 2012
{\em Phys. Rev. Lett.} {\bf 108} 256810

\bibitem{honolka12}
Honolka J, Khajetoorians A A, Sessi V, Wehling T O, Stepanow S, Mi J-L, Iversen B B, Schlenk T,
Wiebe J, Brookes N, Lichtenstein A I, Hofmann P, Kern K and Wiesendanger R 2011
{\em preprint} arXiv:1112.4621


\bibitem{sbv99}
Sachdev S, Buragohain S and Vojta M 1999
{\em Science} {\bf 286} 2479

\bibitem{sv03}
Sachdev S and Vojta M 2003
{\em Phys. Rev. B} {\bf 68} 064419

\bibitem{sandvik07}
H\"oglund K H and Sandvik A W 2007
{\em Phys. Rev. Lett.} {\bf 99} 027205

\bibitem{ribeiro12}
Ribeiro P and Lee P A 2011
{\em Phys. Rev. B} {\bf 83} 235119

\bibitem{kim08}
Kim K-S and Kim M D 2008
{\em J. Phys. Cond. Matter} {\bf 20} 125206

\bibitem{dhochak10}
Dhochak K, Shankar R and Tripathi V 2010
{\em Phys. Rev. Lett.} {\bf 105} 117201


\end{thebibliography}
\end{document}